\documentclass[twocolumn,amssymb,amsmath,floats,showpacs,pre]{revtex4-1}
\usepackage{amsmath,amssymb,bm}
\usepackage{graphicx}
\usepackage{xcolor}
\usepackage[toc,page]{appendix}
\usepackage{comment}
\usepackage[unicode=true]{hyperref}
\usepackage{enumerate}
\graphicspath{{images/}}

\begin{document}

\title{Emergence of protective behaviour under different risk
	perceptions to disease spreading}

\author{
	Mozhgan Khanjanianpak$^{1}$, Nahid Azimi-Tafreshi$^{1}$, Alex Arenas$^{2}$, Jes\'us G\'omez-Garde\~nes$^{3,4}$}

\address{$^{1}$Physics Department, Institute for Advanced Studies in Basic Sciences,  Zanjan 45137-66731, Iran\\
	$^{2}$Departament d'Enginyeria Inform\'atica i Matem\'atiques, Universitat Rovira i Virgili, 430007 Tarragona, Spain\\
	$^{3}$ Department of Condensed Matter Physics, University of Zaragoza, E-50009 Zaragoza, Spain\\
	$^{4}$GOTHAM Lab -- BIFI, University of Zaragoza, E-50018 Zaragoza, Spain}

\begin{abstract}
The behaviour of individuals is a main actor in the
control of the spread of a communicable disease
and, in turn, the spread of an infectious disease
can trigger behavioural changes in a population.
Here, we study the emergence of the individuals’
protective behaviours in response to the spread
of a disease by considering two different social
attitudes within the same population: concerned and
risky. Generally speaking, concerned individuals have
a larger risk aversion than risky individuals. To
study the emergence of protective behaviours, we
couple, to the epidemic evolution of a susceptible-infected-susceptible model, a decision game based
on the perceived risk of infection. Using this
framework, we find the effect of the protection
strategy on the epidemic threshold for each of
the two subpopulations (concerned and risky), and
study under which conditions risky individuals are
persuaded to protect themselves or, on the contrary,
can take advantage of a herd immunity by remaining
healthy without protecting themselves, thanks to the
shield provided by concerned individuals.
\end{abstract}

\maketitle

\section{Introduction}
The prevention and control of the spread of communicable diseases have always constituted a fundamental challenge in human societies~\cite{Anderson,Hethcote,Keeling,Newman}. Nowadays, the Covid-19 pandemic has spread all over the world and, as of October 2021, SARS-CoV-2 has infected more than 244 million people, causing around 5 million deaths. Typically, vaccination is one of the most important preventive measures to prevent or reduce virus propagation and, when its availability is limited, one of the most important issues to study is the effectiveness of different kinds of vaccination strategies aimed at cutting off potential chains of transmission and avoiding as many potential deaths as possible~\cite{Target,Wang,Acquaintance,Qingpeng,Gardon,Zaleta,Peng,Long,Chen,Mozhgan}. However, when vaccines are unavailable or scarce, it is social behavior attaching to preventive measures, the most effective way to reduce and control the spread of the disease~\cite{Fuminori, gosak1, gosak2}. Such preventive strategies include quarantine, self-isolation, social distancing, and the use of prophylactic tools such as face masks. However, most of these behavioral changes and protective measures have associated social and economic costs. Therefore, when their use or application is not subject to law enforcement, regulatory policies or economic support, their application highly depends on an individual decision process. In this scenario, the dynamics of disease spread and the dynamics of individual decision making must be considered as two processes coevolving simultaneously.

From a game theoretical point of view, it seems that behavioural adaptation could be coherently formulated as a game on a complex social network to better understand the evolution of human decisions in response to a given risk, here the spread of a pathogen~\cite{Fredrik}. Several studies have applied game theoretical frameworks coupled to epidemic spreading on populations wherein each rational individual tries to maximize his own payoff according to a self evaluation of the cost-benefit ratio associated to protective measures~\cite{Bauch1,Bauch2,Zhen, Sheryl, Tanimoto}. Although the first studies on this line constitute a combination of epidemic dynamics with a static game theory~\cite{Martial}, the evolutionary game benchmark provides a better description, since it allows agents to update their strategies by evaluating the most successful one according to the current epidemiological state~\cite{Brevan,Zhang,Perisic,Timothy, Wells,Alessio,Steinegger}. This framework appears to be the most appropriate when individual decisions evolve in parallel with the development of an epidemic, allowing each individual to adopt the protection strategy that provides the greatest reward based on the perceived health risk~\cite{Marco, Benjamin}.

Most of the studies devoted to the adoption of prophylactic measures coupled to the unfolding of an epidemic implicitly assume a homogeneous perception of the risk of contagion throughout the population. However, people act differently, reacting heterogeneously to the same information, as evidenced during the current COVID-19 pandemic~\cite{Bavel}. For example, some individuals are more concerned about the evolution of the incidence of cases, trying to be as up to date as possible to take action in an appropriate and responsible manner. On the contrary, others behave in a careless and unconcerned manner or, in the worst case, are critical and suspicious of the alarm messages issued by the health authorities. In~\cite{Mark,Zhishuang}, the authors have studied the effect of these two sub-populations on the spread of infectious diseases. They assume that the behavioral attitude of these two groups can be changed during the epidemic and find  conditions under which the behavioral attitudes can mitigate the disease outbreak.

In this paper, we consider the epidemic model based on risk perception introduced in~\cite{Benjamin} and generalize it to the case when individual reactions to the risk of infection and its associated cost are heterogeneous. In~\cite{Benjamin}, the interaction between individual decisions based on the perception of epidemic risk and the spread of the infectious disease was shown to lead to sustained oscillations over time in the degree of protection adopted by the community. In this study we extend the model by including two different subpopulations: those who are concerned and those who are unconcerned/risky. Thus, faced with the same level of alarm, the first group perceives a much greater risk than the second and is therefore much more likely to take protective measures. To perform our study in a controlled manner, we assume that the partition between subpopulations is constant over time and introduce a parameter that controls the degree of diversity between the two attitudes to the same epidemiological situation. In this way, we show that risky agents start to be protected if the cost of contracting the disease, and the degree of diversity between the two attitudes, exceed certain threshold values. We also show how the fraction of concerned agents can affect the protective and epidemic thresholds of each subpopulation. Finally, we show that herd immunity of risky agents is provided if the number of concerned agents is above a threshold.

The article is organized as follows. In section~\ref{secmodel}, we define our model. In the framework of the Microscopic Markov Chain Approach (MMCA), we derive the dynamic equations of the model. In section~3 we present the main results of our work. In section~3\ref{secEqual} assuming that the fraction of concerned and risky individuals are equal, we find the phase diagrams of the model for the protected and infected fractions of each subpopulation. These results are generalized for different fractions of concerned and unconcerned individuals in section~3\ref{secdiff}, and also the particular case in which risky agents never take protective measures (thus behaving like epidemic deniers) referred as the zealot's limit, is analyzed in section~3\ref{secStub}. The article is concluded in section~\ref{secconclu}.

\section{\label{secmodel}The model}
We consider a population of $N$ individuals connected in pairs and forming a complex network. The infectious pathogen spreads from individual to individual following the connections defined by the social graph and according to the dynamical rules of the SIS (Susceptible-Infected-Susceptible) epidemic model, in which agents can be susceptible (S) or infected (I).  Thus, a susceptible individual at time $t$ becomes infectious from an infectious neighbor with probability $\lambda$ while an infectious agent becomes susceptible with probability $\mu$.
\begin{figure}[t!]
	\centering
	\includegraphics[width=.9\columnwidth]{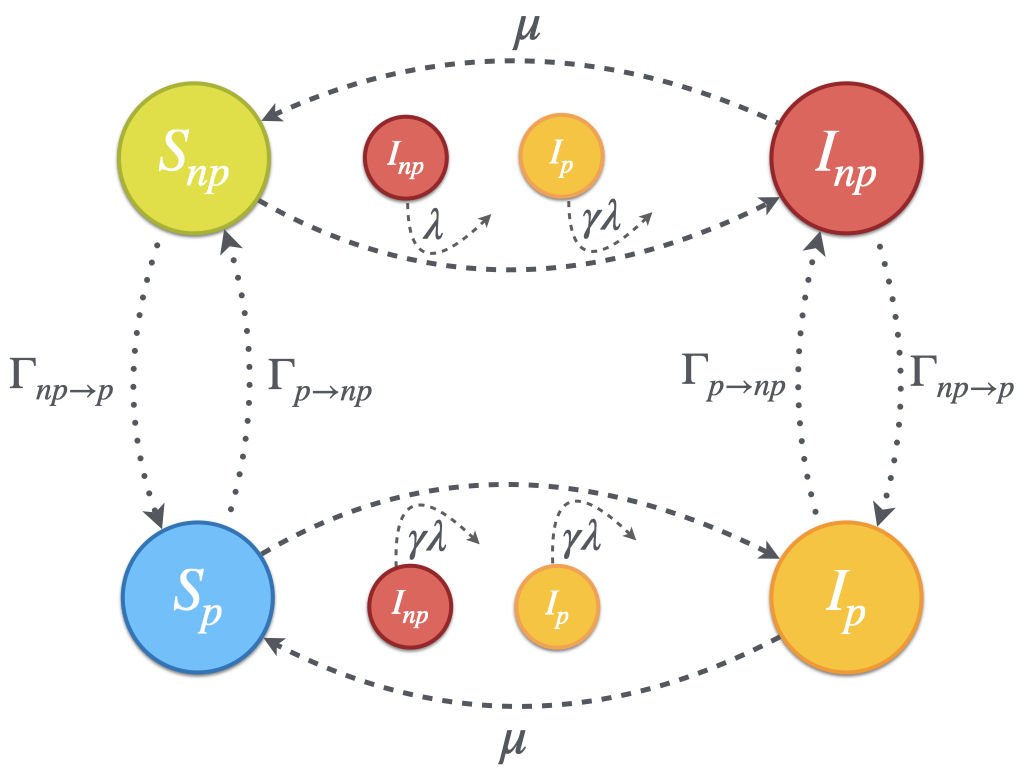}
	\caption{Schematic representation of model dynamics with given transition probabilities. Dashed lines are related to recovery and infection processes while decision-making processes (NP$\to$P and P$\to$NP transitions) are represented by dotted lines.}
	\label{model}
\end{figure}
The classical SIS epidemic model is modified to incorporate the possibility that some agents decide to adopt a prophylactic measure. In particular, we are interested in intermediate measures, such as the use of a mask, that offer partial protection against infection. To this aim, we consider a parameter, $\gamma\in[0,1]$, that determines the decrease in the probability of contagion, which is transformed into $\gamma\lambda$ when in the encounter between an infectious agent and a susceptible one of the two adopts the prophylactic strategy. Thus, $\gamma=0$ would imply that the preventive measure is perfect while $\gamma=1$ means that the measure is completely useless. Similar to~\cite{Benjamin}, we assume that the preventive mechanism is not increased by the bilateral use of protective measures, and thus the reduction in infectivity is linear, $\gamma\lambda$, rather than quadratic, $\gamma^2\lambda$, in the case where both individuals (infectious and susceptible) are protected.

The use of the prophylactic measure depends on the individual free-will. To address the dynamics associated to this individual choice we couple, as in~\cite{Benjamin}, the SIS epidemic model to a two-strategy Protected-Unprotected ($P$-$NP$) decision game in which agents decide whether or not to adopt a prophylactic measure based on the associated costs and the disease incidence for each strategy. In particular, the choice of whether or not to take the prophylactic measure carries an associated cost and individuals must therefore assess the appropriateness of its use based on their perception of the risk of contagion.  To account for the decision process, agents evaluate the strategic choice ($P$ or $NP$) based on: (i) the cost of contracting the disease ($T$), (ii) the cost of the prophylactic measure ($c$), and (iii) the risk of contracting the disease, measured through its incidence (i.e. the fraction of infectious individuals $I$). The evaluation of the contagion risk at each time step is used to compute the expected benefit of each strategy, the so-called payoffs, $P_p$ and $P_{np}$, that in their turn are used to construct the probabilities that drive strategic changes  $\Gamma_{p\rightarrow np}$ and $\Gamma_{np\rightarrow p}$. These probabilities change in time according to the epidemic incidence and thus govern the evolution of the strategic partition of the population. The strategic partition strongly influences the epidemic evolution, being the spreading of the pathogen favored when $NP$ strategy dominates or being its propagation mitigated when prophylaxis is widely adopted. A schematic plot of the coevolution of the spreading and decision processes is shown in Fig.~\ref{model}.

Finally, the main novelty of this work lies in dividing the population into two groups: Concerned (C) and Risky/unconcerned (R). These groups have a different perception of risk and, therefore, under the same epidemic scenario show different behavioral responses. For example, concerned agents have a higher perception of the cost of contracting the disease than risky individuals ($T^C>T^R$) while the cost associated with protection is perceived as lower by the former than by the latter ($c^R>c^C$). Thus, on one hand, concerned individuals perceive a higher risk associated with the disease and a lower cost of protection so, consequently, they are inclined to adopt prophylactic measures. On the other hand, risk-takers are more relaxed in the face of epidemic alarm and are more likely to take risky decisions. In this paper we assume that the populations of concerned and risky agents remain constant during the dynamics, i.e. neglecting possible changes in risk perception during the course of the epidemic dynamics. In particular we fix a parameter $f$, the fraction of concerned individuals in the population so that the number of concerned and risky agents are $N_C=fN$ and $N_R=(1-f)N$.

\begin{figure}[t!]
	\centering
	\includegraphics[width=1.0\columnwidth]{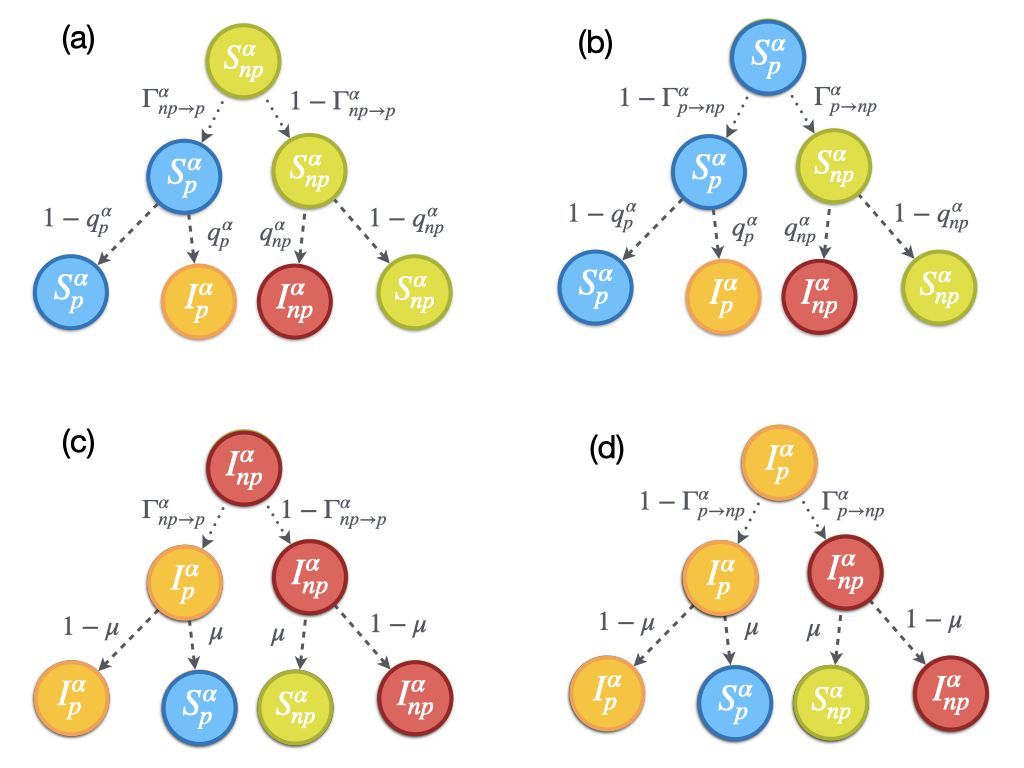}
	\caption{Transition probability trees for the states $X_y^{i,\alpha}(t)$ at each time step. The root shows initial state at $t$ and the leaves represent the states at the next time $t+1$. The first-row arrows show the P$\to$NP and NP$\to$P processes with probabilities $\Gamma^{i,\alpha} _{p \to np}$ and $\Gamma^{i,\alpha} _{np \to p}$. The second-row arrows denote probabilities $\mu$, $q^{i,\alpha}_{np}$ or $q^{i,\alpha}_p$ governing the changes in the epidemiological states of each node. }
	\label{mmc}
\end{figure}

\subsection{Markovian formulation of the model}
To perform the model analysis we will translate the above basic evolution rules into discrete-time Markovian equations~\cite{GOMEZ}. For each individual there are four accessible states (compartments): $S_p$, $S_{np}$, $I_p$ and $I_{np}$. Therefore, the state at time $t$ of an individual $i$ ($i=1,...,N$) belonging to group $\alpha$ ($\alpha\in\{C,R\}$) is defined by the probabilities  $X_y^{i,\alpha}(t)$, with $X\in\{S,I\}$ and $y\in\{p,np\}$ that agent $i$ is in one of the former four states. Obviously, for a given agent $i$ belonging to group $\alpha$ these probabilities satisfy the normalization condition:
\begin{equation}
	S^{i,\alpha}_p(t)+S^{i,\alpha}_{np}(t)+I^{i,\alpha}_p(t)+I^{i,\alpha}_{np}(t)=1\;.
\end{equation}

At each time step, each agent determines her state according to the values of the probabilities $\{X_y^{i,\alpha}(t)\}$ that are updated according to the possible transitions among the four compartments as plotted in Fig.~\ref{mmc}. In particular, the Markovian equations governing the evolution of the four probabilities associated to a given individual $i$ belonging to group $\alpha$ are:
\begin{widetext}
\begin{align}
	S^{i,\alpha}_p (t+1) =& (1-\Gamma^\alpha _{p \to np}(t)) \Big{[} S^{i,\alpha}_p(t)(1-q^{i,\alpha}_p(t))
	+ \mu I^{i,\alpha}_p(t) \Big{]} + \Gamma^\alpha _{np \to p}(t) \Big{[} S^{i,\alpha}_{np}(t)(1-q^{i,\alpha}_p(t)) + \mu I^{i,\alpha}_{np}(t)\Big{]}
	\label{Sp}
\end{align}
\begin{align}	
	S^{i,\alpha}_{np} (t+1) =& \Gamma^\alpha _{p \to np}(t) \Big{[} S^{i,\alpha}_p(t)(1-q^{i,\alpha}_{np}(t))
	+ \mu I^{i,\alpha}_p(t)\Big{]}	+(1- \Gamma^\alpha _{np \to p} (t))\Big{[}S^{i,\alpha}_{np}(t)(1-q^{i,\alpha}_{np}(t)) + \mu I^{i,\alpha}_{np}(t)\Big{]}
	\label{Snp}	
\end{align}
\begin{align}
	I^{i,\alpha}_p (t+1) =& (1-\Gamma^\alpha _{p \to np}(t)) \Big{[} S^{i,\alpha}_p(t)q^{i,\alpha}_p(t)
	+ (1-\mu)I^{i,\alpha}_p(t)\Big{]} + \Gamma^\alpha _{np \to p}(t)\Big{[}S^{i,\alpha}_{np}(t)q^{i,\alpha}_p(t) + (1-\mu)I^{i,\alpha}_{np}(t)\Big{]}
	\label{Ip}
\end{align}
\begin{align}	
	I^{i,\alpha}_{np}(t+1) =& \Gamma^\alpha _{p \to np}(t) \Big{[} S^{i,\alpha}_p(t)q^{i,\alpha}_{np}(t)
	+ (1-\mu)I^{i,\alpha}_p(t) \Big{]} +(1- \Gamma^\alpha _{np \to p}(t)) \Big{[} S^{i,\alpha}_{np}(t)q^{i,\alpha}_{np}(t) + (1-\mu)I^{i,\alpha}_{np}(t)\Big{]}
	\label{Inp}	
\end{align}
\end{widetext}
where $q^{i,\alpha}_p(t)$ ($q^{i,\alpha}_{np}(t)$) indicates the probability that agent $i$ of type $\alpha$, being susceptible and protected (non-protected) at time $t$, is infected at this time step. Also, $\Gamma^{\alpha}_{p\rightarrow np}(t)$ ($\Gamma^{\alpha}_{np\rightarrow p}(t)$) is the probability that an agent that is protected (non-protected) at time $t$ decides to update the strategy to non-protected (protected) at this time step. Probabilities $q^{i,\alpha}_p(t)$, $q^{i,\alpha}_{np}(t)$, $\Gamma^{\alpha}_{p\rightarrow np}(t)$ and $\Gamma^{\alpha}_{np\rightarrow p}(t)$ are derived below.

\begin{figure*}[t!]
	\centering
	\includegraphics[width=1.3\columnwidth]{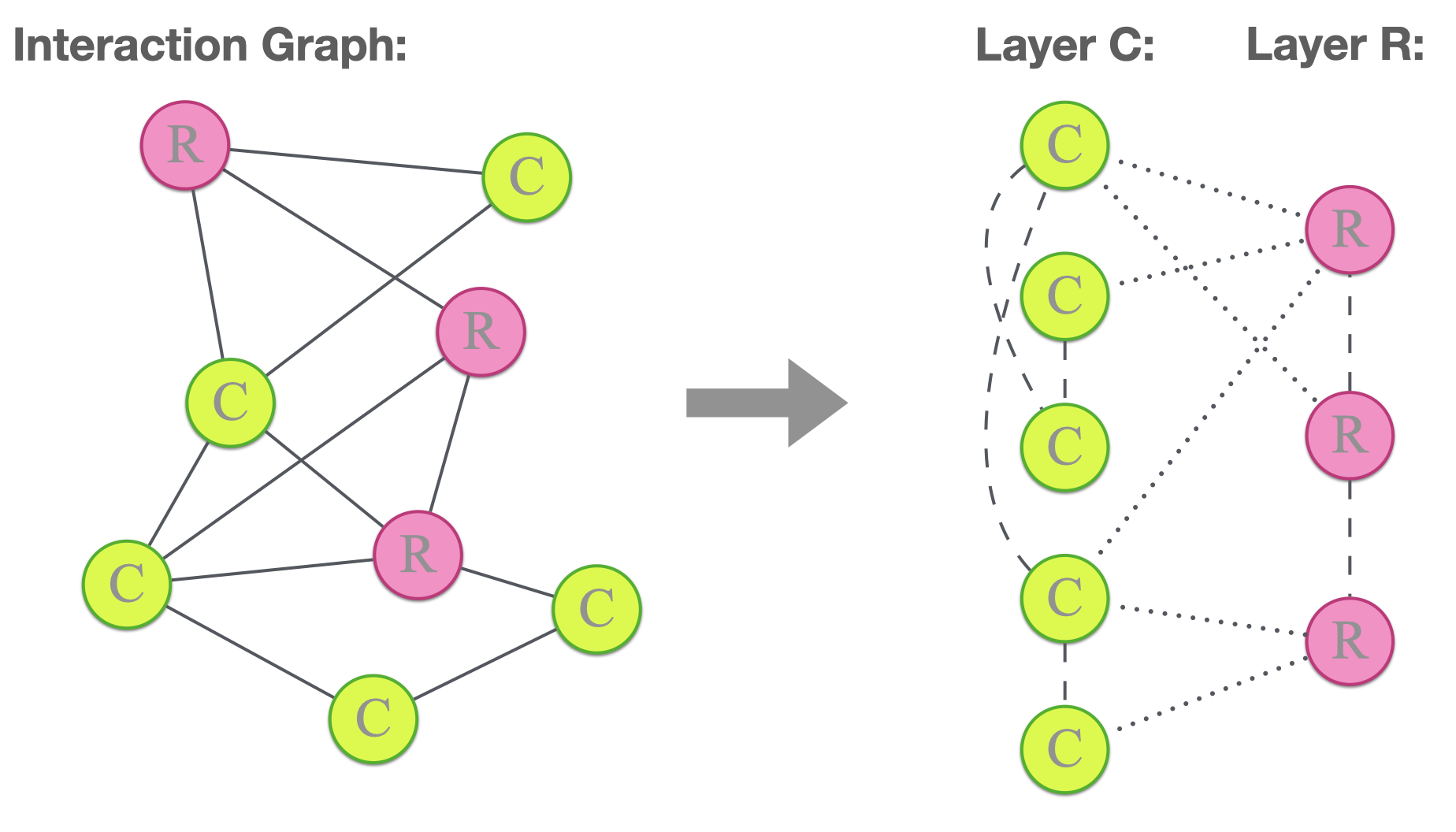}
	\caption{(a) The network consists of eight nodes. Green nodes indicate the concerned (C) agents and red ones are risky (R). (b) A two-layer representation of the model. Concerned (Risky) nodes belong to layer~C (layer~R), connected through intra-layer links, while the risky and concerned nodes are connected through inter-layer links.
		Intra-layer and inter-layer links are represented by dashed and dotted lines respectively.}
	\label{multilayer}
\end{figure*}
\subsubsection{Infection probabilities}
To derive the probabilities $q^{i,\alpha}_p(t)$ and $q^{i,\alpha}_{np}(t)$, let us interpret the interaction network as a two-layer graph, so that concerned and risky nodes belong to each of the two layers respectively. This bilayer network is composed of $N_{C}=fN$ and $N_R=(1-f)N$ nodes in each layer and contains intra-layer ($R-R$ and $C-C$) and inter-layer ($R-C$) links as shown in Fig.~\ref{multilayer}. Obviously, the number of links of each type is constant, since we assume that the fraction of concerned ($f$) and risky ($1-f$) agents is constant.

In this bilayer structure, a susceptible node in layer $\alpha$, can become infected through one of its infected neighbors in the same layer or through an inter-layer link connecting to an infected agent. Following the epidemic dynamics $S\leftrightharpoons I$ described above, a not-protected susceptible agent, i.e. having a state $S^\alpha_{np}(t)$, can contract the disease from each of those $I^\alpha_{np}(t)$ neighbors in the both layers with probability $\lambda$,  while infections coming from protected infected neighbors, $I^\alpha_{p}(t)$, in both layers occur with probability $\gamma \lambda$. However a protected susceptible agent, $S^\alpha_{p}(t)$, will be infected with probability $\gamma \lambda$, regardless the contact occur through an intra-layer or an inter-layer link with agents in states $I^\alpha_{p}(t)$ and $I^\alpha_{np}(t)$. Consequently, we can write the following equations:
\begin{widetext}
\begin{align}
	q^{i,\alpha}_p(t) = 1 - \displaystyle\prod_{j=1}^{N}  \big{[} 1 - \lambda \gamma A_{ij}^{\alpha\alpha} ( I^{j,\alpha}_p (t) + I^{j,\alpha}_{np}(t) )  \big{]} \displaystyle\prod_{j=1}^{N}  \big{[} 1 - \lambda \gamma A_{ij}^{\alpha\beta} ( I^{j,\beta}_p (t) + I^{j,\beta}_{np} (t) )  \big{],}
	\label{qp}
\end{align}
\begin{align}
	q^{i,\alpha}_{np}(t) = 1 -\displaystyle\prod_{j=1}^{N} \big{[} 1 - \lambda A_{ij}^{\alpha\alpha} (\gamma I^{j,\alpha}_p(t) + I^{j,\alpha}_{np} (t) )\big{]} \displaystyle\prod_{j=1}^{N} \big{[} 1 - \lambda  A_{ij}^{\alpha\beta} ( \gamma I^{j,\beta}_p (t) + I^{j,\beta}_{np} (t) ) \big{]}
	\label{qnp}
\end{align}
\end{widetext}
where $\alpha , \beta \in \{C,R\}$, and ${\bf A}^{\alpha\beta}$ denotes an $N\times N$ adjacency matrix whose element $(i,j)$ equals to 1 if there is a connection between agent $i$ of type $\alpha$ and agent $j$ of type $\beta$, while $A_{ij}^{\alpha\beta}=0$ otherwise. Note that the overall adjacency matrix of the network ${\bf A}$ can be written as ${\bf A}={\bf A}^{CC}+{\bf A}^{RR}+{\bf A}^{CR}$. Thus, the former equations can be read as $1$ minus the probability of not being infected by any infectious agent. In both expressions, this latter probability is split in two terms: the first product of brackets gives the probability of not being infected via intra-layer ($R-R$ or $C-C$) links, while the second product indicates the probability of not contracting the virus from an agent of different type, i.e., through $R-C$ links.

\subsubsection{Strategic update probabilities}
To round off the derivation of the Markovian equations we now show the expression for the probabilities associated to the strategic updates, $\Gamma^{\alpha}_{p\to np}(t)$ and $\Gamma^\alpha_{np\to p}(t)$, for each type of agent $\alpha\in\{C,R\}$. To do so, we first assign the payoff associated to each strategy at a given time $t$ for an individual of class $\alpha$:
\begin{align}
	P_p^\alpha(t) = -c^\alpha - T^\alpha \frac{I_p(t)}{I_p(t)+S_p(t)}
	\label{Pp}
\end{align}
\begin{align}
	P_{np}^\alpha(t) = - T^\alpha \frac{I_{np}(t)}{I_{np}(t)+S_{np}(t)}
	\label{Pnp}
\end{align}
where 
\begin{eqnarray}
	S_p(t)&=&N^{-1}\sum_{\alpha=\{C,R\}}^{}\sum_{i=1}^{N_\alpha}S_p^{i,\alpha}(t)\;,\\
	S_{np}(t)&=&N^{-1}\sum_{\alpha=\{C,R\}}^{}\sum_{i=1}^{N_\alpha}S_{np}^{i,\alpha}(t)\;,\\
	I_p(t)&=&N^{-1}\sum_{\alpha=\{C,R\}}^{}\sum_{i=1}^{N_\alpha}I_p^{i,\alpha}(t)\;,\\
	I_{np}(t)&=&N^{-1}\sum_{\alpha=\{C,R\}}^{}\sum_{i=1}^{N_\alpha}I_{np}^{i,\alpha}(t)\;,
\end{eqnarray}
are the expected (average) fraction of individuals in $S_p$, $S_{np}$, $I_p$ and $I_{np}$ compartments respectively. As introduced above parameters $c^\alpha$ and $T^\alpha$ denote respectively the costs associated to protection and infection for an agent of class $\alpha$. According to the behavioral responses of the concerned and risky agents, we assume that $c^{R} > c^{C}$ and $T^{C} > T^{R}$. In particular, we consider the relation between parameters $c$ and $T$ for the concerned and risky people as follows:
\begin{equation}
	T^{R}= \delta T^{C}
	\label{TR}
\end{equation}
and
\begin{equation}
	c^{C} = \delta c^{R}
	\label{CC}
\end{equation}
where $\delta \in [0,1]$, is a parameter that accounts for the awareness gap between concerned and risky agents.

Finally, once agents have estimated the payoff associated to each strategy they can choose between acting as protected or not-protected in the next time step. In particular, for an agent of type $\alpha$, $\Gamma^\alpha_{p\to np}(t)$ and $\Gamma^\alpha_{np\to p}(t)$ denote the update probabilities for the strategic choice $P\leftrightharpoons NP$. Here we take the usual discrete-time and finite-population analogue of the replicator evolutionary rule \cite{Hauert04,Roca,Vilone} for the form of these update probabilities:
\begin{equation}
	\Gamma^\alpha_{p\to np}(t) = \frac{P^\alpha_{np}(t) - P^\alpha_p(t)}{c^\alpha + T^\alpha} \Theta(P^\alpha_{np}(t) - P^\alpha_p(t))
	\label{Gp}
\end{equation}
\begin{equation}
	\Gamma^\alpha_{np\to p}(t) = \frac{P^\alpha_{p}(t) - P^\alpha_{np}(t)}{c^\alpha + T^\alpha} \Theta(P^\alpha_{p}(t) - P^\alpha_{np}(t))
	\label{Gnp}
\end{equation}
where $\Theta(x)$ is the Heaviside function:
\begin{equation}
	\Theta(x)=\left\{ \begin{array}{l}
		{1 \qquad x\geq 0} \\
		{0 \qquad x< 0.}
	\end{array} \right.
	\label{SF}
\end{equation}

\section{Results} 

Once we have introduced the dynamical rules of the model and the associated Markovian equations, we proceed to analyze the behavior arising from the interplay between the SIS spreading dynamics, the evolutionary dynamics for individual decisions, and the partition of the population between concerned and risky agents. We do so by iterating numerically Eqs.~(\ref{Sp})-(\ref{Inp}) from a given initial condition to construct the trajectory of the probabilities associated to each of the agents until a (static or dynamical) equilibrium is reached. All the results presented here consider a simplified network of social interactions represented by an Erd\"os-R\'enyi (ER) graph with size $N=2000$ and mean degree $\langle k \rangle= 10$. Although this is a very naive approximation to the structure of real social networks, it represents a starting point to understand the physical properties of the critical system.

Our main interest is to study how contagion and the use of protective measures co-evolve in the groups of concerned and risky individuals so to gain insights about the mutual influence between these antagonistic behaviors in a given population. To this aim we should define the following observables of interest: the relative fraction of risky agents that are infected ($I^R$), the relative fraction of risky agents that are protected ($P^R$), the relative fraction of concerned agents that are infected ($I^C$), and the relative fraction of concerned agents that are protected ($P^C$).
Starting from the microscopic probabilities $X_{y}^{i,\alpha}(t)$ associated to each individual it is easy to compute the average (expected) occupation at a given time $t$ of the $4$ possible states accessible to each group $\alpha$ as: $X_{y}^{\alpha}(t)=N^{-1}\sum_{i=1}^{N_{\alpha}}X_{y}^{i,\alpha}(t)$. The corresponding four average quantities fulfil, at any time, the following conservation laws:
\begin{align}\label{f}
	S^C_p(t) + S^C_{np}(t) + I^C_p(t) + I^C_{np}(t)=f,
\end{align}
\begin{align}\label{1-f}
	S^R_p(t) + S^R_{np}(t) + I^R_p(t) + I^R_{np}(t) =(1-f).
\end{align}
Thus, it is possible to define the relative fractions of interest to our study as:
\begin{align}
	I^\alpha=\frac{N(I_{np}^\alpha + I_p^\alpha)}{N_{\alpha}},
	\label{IR}
\end{align}
\begin{align}
	P^\alpha=\frac{N(I_{p}^\alpha + S_p^{\alpha})}{N_{\alpha}}.
	\label{PR}
\end{align}


\begin{figure}[t!]
	\centering
	\includegraphics[width=1.0\columnwidth]{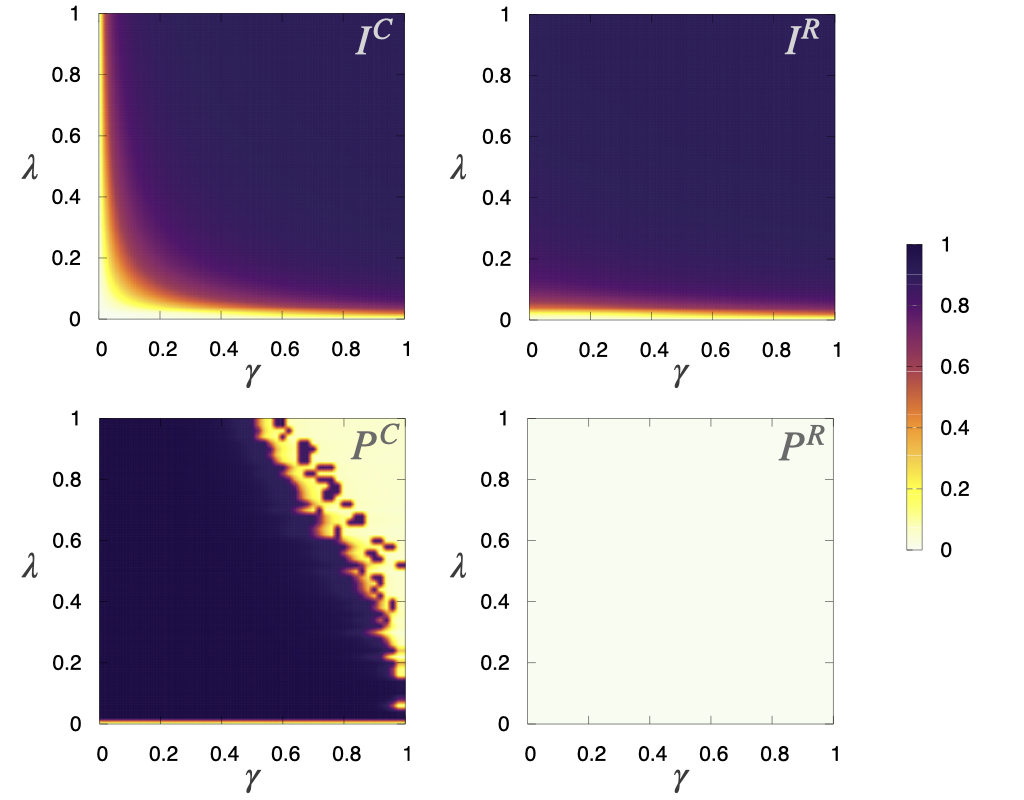}
	\caption{From these diagrams it is clear that the healthy phase (according to the color bar, white=0 corresponding 0\% infected) is much larger for the concerned group, specially for small values of
$\gamma$. In both populations the transition from the disease-free regime and the epidemic state is smooth. Panels in the bottom show, respectively, the fraction (according to the color bar, dark blue=1 corresponding 100\% protected) of concerned and risky agents that are protected ($P^C$ and $P^R$) as a function of $\gamma$ and $\lambda$. The diagrams correspond to an ER network with $N=2000$ nodes and mean degree of $\langle k \rangle=10$, where $f=0.5$, $T^{C}=10, \delta=0.01, \mu=0.1$ and $c^{R}=1$.}
	\label{default}
\end{figure}

\subsection{Identical subpopulation sizes ($f=1/2$)}
\label{secEqual}

Let us first consider the case $f=1/2$, so that the number of concerned and risky individuals in the population is equal ($N_C=N_R=N/2$), and analyze the behavior of $I^{\alpha}$ and $P^{\alpha}$ as a function of the contagion probability, $\lambda$, and the failure probability of prophylaxis, $\gamma$.
In Fig.~\ref{default} we show the phase diagram for these quantities in the parameter plane ($\gamma,\lambda$) and fix the values of the remaining parameters as reported in the caption of Fig.~\ref{default}. For the concerned group, different regimes similar to those found in~\cite{Benjamin} are obtained. Namely:
\begin{itemize}
	\item {\em Healthy state}: wherein all concerned agents are susceptible but not protected. This happens for $\lambda<\lambda_c$, i.e., for contagion probabilities below the epidemic threshold.
	\item {\em Healthy-Protected state}: in which all concerned agents are susceptible and protected. This occurs for $\lambda<\lambda_c$ and $\gamma\ll 1$.
	\item {\em Infected-Protected state}: wherein a large fraction of concerned individuals get infected but, nevertheless, they adopt protection.
	\item {\em Infected-Not-protected state}: in which as the protection effectiveness decreases ($\gamma$ increases), for large values of infection probability $\lambda$, concerned agents cease to adopt protection. The curve $\lambda(\gamma)$ that shapes the border between the Infected-Protected and the Infected-Not-Protected regimes defines the so-called protection threshold~\cite{Benjamin}.
\end{itemize}

In its turn, the behaviour of risky agents is totally different when, as in Fig.~\ref{default}, the value of $\delta$ is very small, so that the awareness gap between concerned and risky people becomes very large. In this case, the phase diagram for the fraction of protected risky individuals $P^R(\gamma,\lambda)$, shows that risky agents refrain from protecting themselves regardless of the value of  $\lambda$ and $\gamma$. Even for high effective protection and low infection probabilities ($\gamma\ll 1$ and $\lambda\gtrsim \lambda_c$), the chosen strategy is always NP. Thus, the phase diagrams correspond to the usual SIS with only two regimes: the disease-free ($\lambda<\lambda_c$) and the epidemic ($\lambda>\lambda_c$) ones.

An interesting question is whether risky agents change their strategy by adopting protection for certain values of parameters. To unveil that, below we study the role of $T^C$ (or equivalently $T^R$), and $\delta$, while keeping $c^R=1$ without loss of generality.

\subsubsection{Role of the contagion cost $T$}
\label{secEquT}
\begin{figure}[b!]
	\includegraphics[scale=0.4]{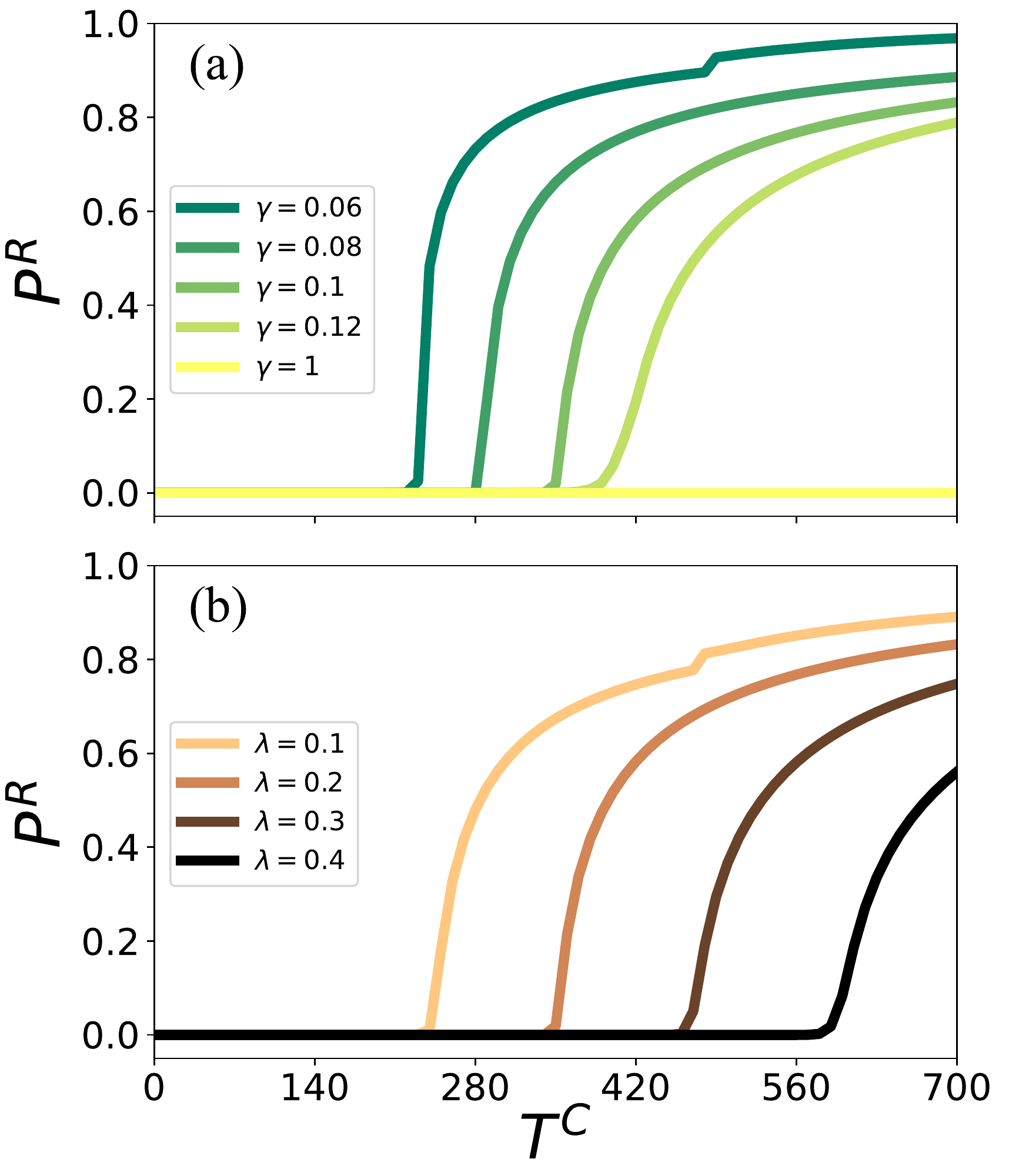}
	\caption{Stationary values for $P^R$ as a function of $T^C$. These diagrams correspond to an ER network with $N=2000$ and $\langle k \rangle=10$, where $f=0.5, \delta=0.01, \mu=0.1$ and $c^{R}=1$. In $(a)$ $ \lambda=0.2$  and in $(b)$ $\gamma=0.1$.}
	\label{Trole}
\end{figure}
When the cost of contracting an infection ($T^C$) increases, even risky individuals start to adopt prophylactic measures as shown in Fig.~\ref{Trole}, showing the evolution of $P^R$ as a function of $T^C$ for different values of the protection effectiveness $\gamma$ and the contagion probability $\lambda$.
\begin{figure*}[t!]
	\centering
	\includegraphics[width=2\columnwidth]{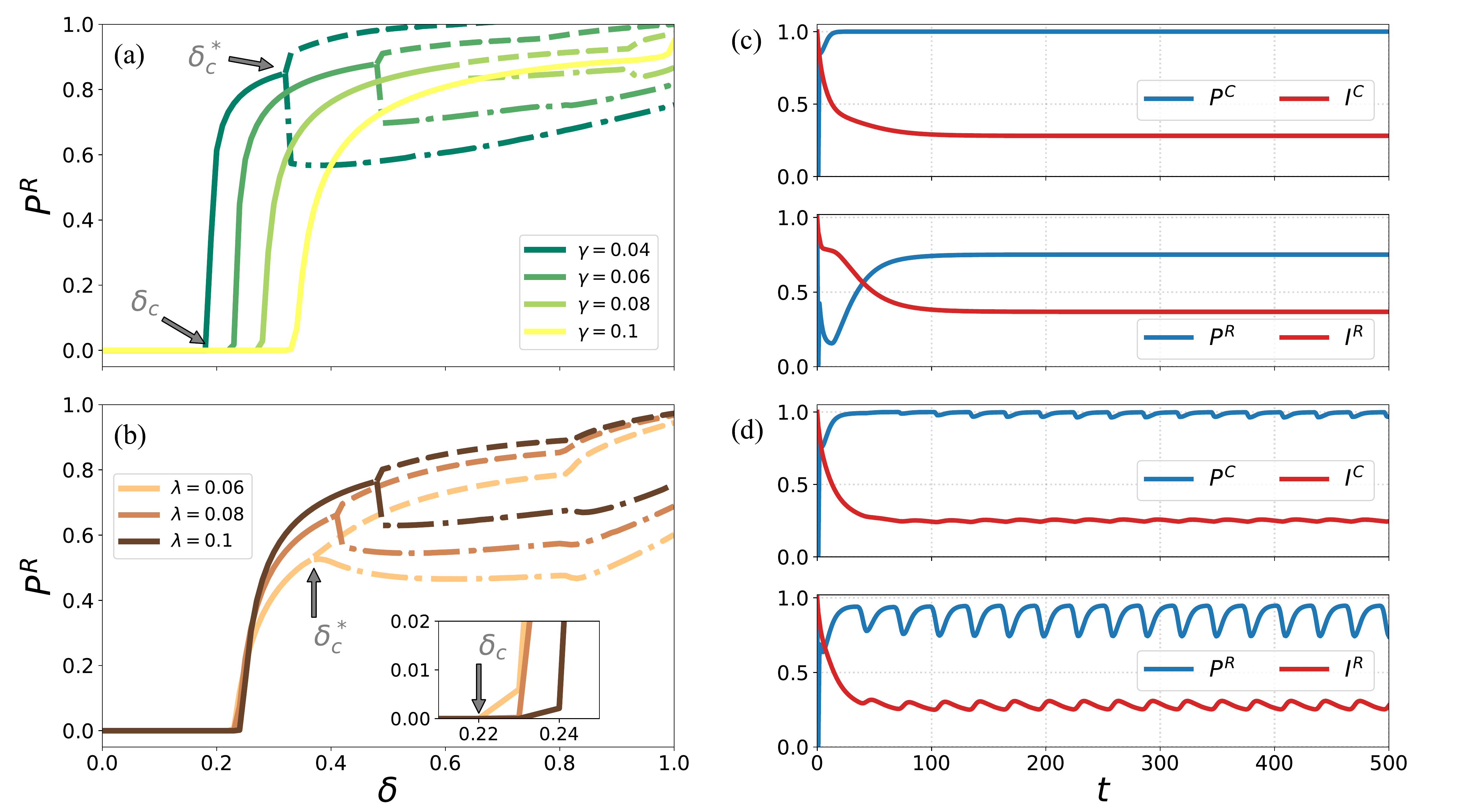}
	\caption{(a)-(b) Bifurcation diagrams for fraction of $P^R$ as a function of $\delta$. The numerical results are obtained for the ER network with $N=2000$ and $\langle k \rangle=10$. Parameters are $T^{C}=10, \mu=0.1$, $c^{R}=1$, $f=0.5$ and
		in $(a)$ $\lambda =0.2$ and in $(b)$ $\gamma =0.1$. Bifurcation points $\delta_c$ and $\delta_c^*$ are shown on the figure. Solid lines denote stable fixed-points while dashed-dotted and dashed lines show the lower and upper turning points of stable limit cycles, respectively. (c)-(d) Numerical results for fraction of protected and infected compartments of both risky and concerned individuals in time $t$ on the ER network with $N=2000$ and $\langle k \rangle=10$. Parameters are $T^{C}=10, \mu=0.1$, $c^{R}=1, \lambda=0.2, \gamma=0.06$, $f=0.5$ and in (c) $\delta =0.3$ and in (d) $\delta =0.8$.}
	\label{limit}
\end{figure*}
In both cases the reported curves $P^R(T^C)$ show a clear phase transition so that above a threshold value, $T^C_c$, risky agents start to take protection measures. The precise value of the cost threshold, $T^C_c$, depends on the values of $\gamma$ and $\lambda$, increasing when the effectiveness of protection decreases and the contagion probability increases.

\subsubsection{Role of the awareness gap}

Up to now, we have considered that the awareness gap between risky and concerned agents is extremely large $\delta=0.01$, thus being very far from an homogeneous response to epidemic risk in the entire population. As $\delta$ grows the awareness gap decreases, recovering the homogenous response limit studied in~\cite{Benjamin} for $\delta=1$.

To monitor the transition between heterogeneous and homogeneous response to risk we have explored the evolution of $P^R$ as a function of $\delta$. Panels (a) and (b) in Fig.~ \ref{limit} show the bifurcation diagram for $P^R$ as a function of $\delta$ for different values of the probability of protection failure $\gamma$ and the infection probability $\lambda$. Again, a phase transition for the adoption of protection by risky agents shows up when $\delta$ equals some threshold value $\delta_c$. Interestingly, the value of $\delta_c$ needed for risky agents to take protective measures notably increases as the measure becomes more inefficient, but it slightly changes with the probability of contagion $\lambda$.

When protection is highly effective, i.e. for small values of $\gamma$, the increase of $\delta$ beyond $\delta_c$ drives the system towards a supercritical Hopf bifurcation point, denoted as $\delta^*_c$, in which the fixed point for the protection levels of risky (and also that of concerned)  individuals loses its stability while a limit cycle in which the fraction $P^R$ (and $P^C$) oscillates in time in a sustained way. Solid lines in Fig.~\ref{limit}, show stable fixed-points for the range of $0 < \delta < \delta_c^*$, while dashed-dotted and dashed lines identify lower and upper turning points of the stable limit cycles, respectively. From these diagrams we observe that the increase of $\gamma$ and $\lambda$ shifts the Hopf bifurcation to larger values of $\delta$, so that for large enough values of $\gamma$ and $\lambda$, the oscillatory behaviour vanishes.

\begin{figure}[b!]
	\centering
	\includegraphics[width=1\columnwidth]{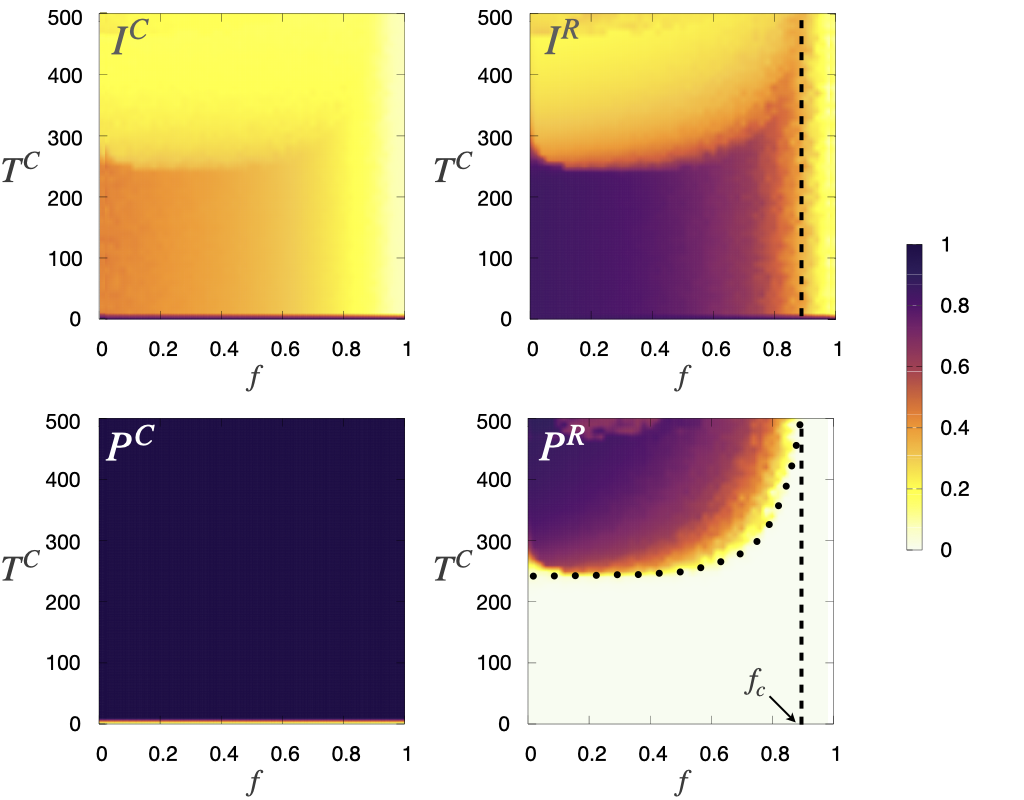}
	\caption{Phase diagrams for the infected and protected fraction of the concerned ($I^C$ and $P^C$) and risky ($I^R$ and $P^R$) agents in the space $T^C-f$. Numerical results are obtained on the ER network with $N=2000$ and $\langle k \rangle=10$. Other parameters are $\lambda=0.1, \delta=0.01, \mu=0.1$, $c^{R}=1$, and $\gamma=0.1$. The dashed line in diagrams for $I^R$ and $P^R$ highlight the estimated value of $f_c$ while the dotted curve is an estimation of the protection threshold.}
	\label{resultsGams}
\end{figure}

Panels (c) and (d) of Fig.~\ref{limit} shows the behavior of both concerned and risky groups before ($\delta_c < \delta < \delta_c^*$) and after ($\delta >\delta_c^*$) the Hopf bifurcation. From these panels we also observe that the amplitude of oscillations is smaller for concerned people than for the risky group. This implies that strategic-changes are more likely for risky agents. For $\delta$ close enough to $1$, behavioral responses of both groups are identical, as it is expected (figure not shown).

\subsection{Unbalanced populations of concerned and risky individuals}\label{secdiff}

Our previous results have shown that when concerned and risky players equally populate the network, risky agents take advantage of the protective effort of concerned ones when the cost associated with the disease does not exceed a threshold, $T^C_c$ or when $\delta$ is less than $\delta_c$. At this point, one would ask whether increasing the fraction of risky agents can lead to a change in their strategy, in particular of those threshold values corresponding to their onset of protection. To this aim, we have studied how contagion and protection patterns change when the fraction $f$ of concerned agents varies.
\begin{figure}[t!]
	\begin{tabular}{c}
		\includegraphics[width=0.9\columnwidth]{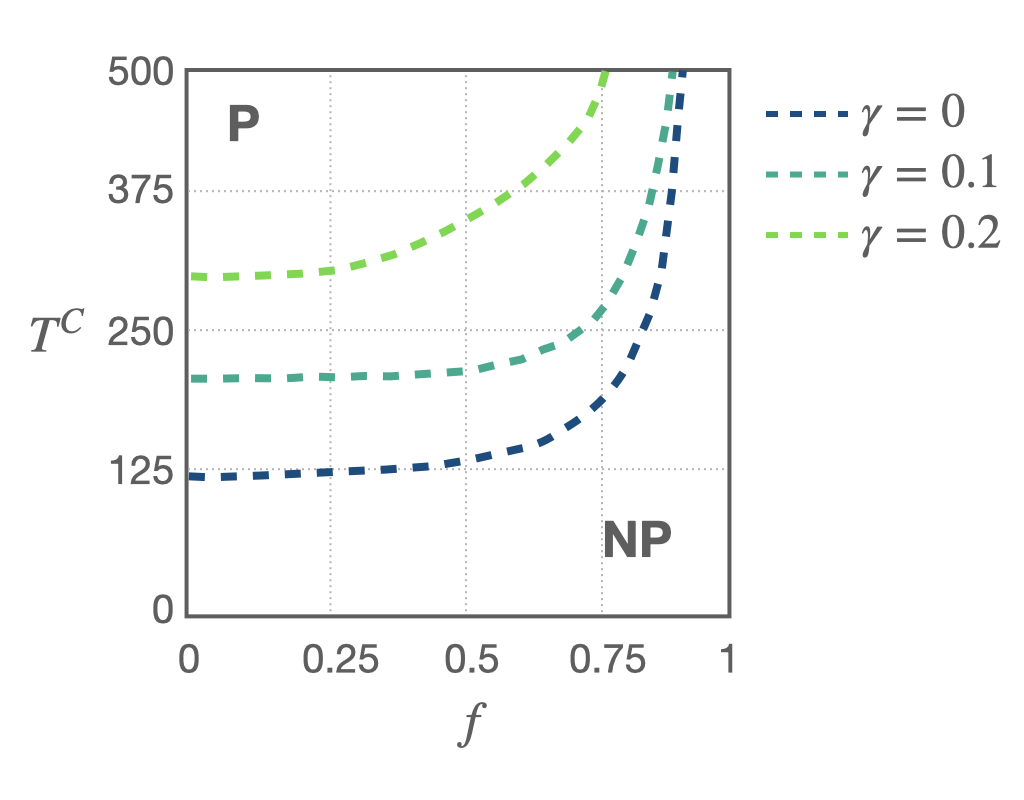}
	\end{tabular}
	\centering
	\caption{Protection threshold curves of the risky agents in $(T^C, f)$ space, for three different values  $\gamma=0, 0.1$ and $0.2$. The region wherein risky agents (do not) protect themselves is labelled as (NP) P which is (bottom) above of each curve.}
	\label{thre}
\end{figure}

In Fig.~\ref{resultsGams} we show the phase diagrams for $P^R$, $I^R$, $P^C$, and $I^C$ as a function of $f$ and $T^C$. While for concerned agents  protection is fully adopted, risky agents show their resistance to protect themselves unless $T^C$ is large enough, $T^C>T^C_c$. However, from panel (d) it becomes clear that the the value of $T^C_c$ remains almost constant for small and intermediate values of $f$ while showing a sharp increase when the fraction of concerned agents $f$ is increased.This effect allows us to define a threshold $f_c$ (white dashed line) so that for a given value of $f>f_c$ risky agents cease to adopt protection regardless of the infection cost $T^C$ . To shed light on this effect we show in Fig.~\ref{thre} the protection threshold curves $T^C_c(f)$, separating phases $P^R=0$ and $P^R>0$ for different values of $\gamma$, showing that the higher the protection efficiency (smaller $\gamma$), the smaller the value of $T^C_c$.

These results seem to indicate that when the fraction of concerned individuals is relatively large, in particular when $f>f_c$, the risky minority can take full advantage of the protective efforts of the concerned to get rid of the contagion without paying any cost.

\subsection{The limit $\delta=0$: Risky and Concerned zealots}\label{secStub}

 Let us now shed light on the characterization of the value $f_c$, $i.e.$ the critical fraction of concerned
 individuals that allow risky ones to free ride on the protective efforts of the former. To this aim we
 consider the limit when $\delta=0$.  In this limit both risky and concerned agents act like zealots: on one hand for risky agents the ratio between the cost of protection and the cost of the disease becomes infinite (since they neglect the cost of contracting the diseases $T^R=0$)  and thus never accept protection, while for concerned agents the former ratio is $0$ (since the cost associated to protection is $c^C=0$). This behavior becomes evident by evaluating  Eqs.~(\ref{Pp}) and (\ref{Pnp}) for  risky agents, for whom the pay-off associated to the NP strategy is always larger than that of the P strategy. Hence, no matter how severe the incidence of the disease is, their decision is never changed. On the other hand, for the concerned agents $c^C$ vanishes in this limit, and only $\frac{I_p(t)}{I_p(t)+S_p(t)}$ and $\frac{I_{np}(t)}{I_{np}(t)+S_{np}(t)}$ fractions affect the protection dilemma for this group.

\begin{figure}[t!]
	\centering
	\includegraphics[width=1.0\columnwidth]{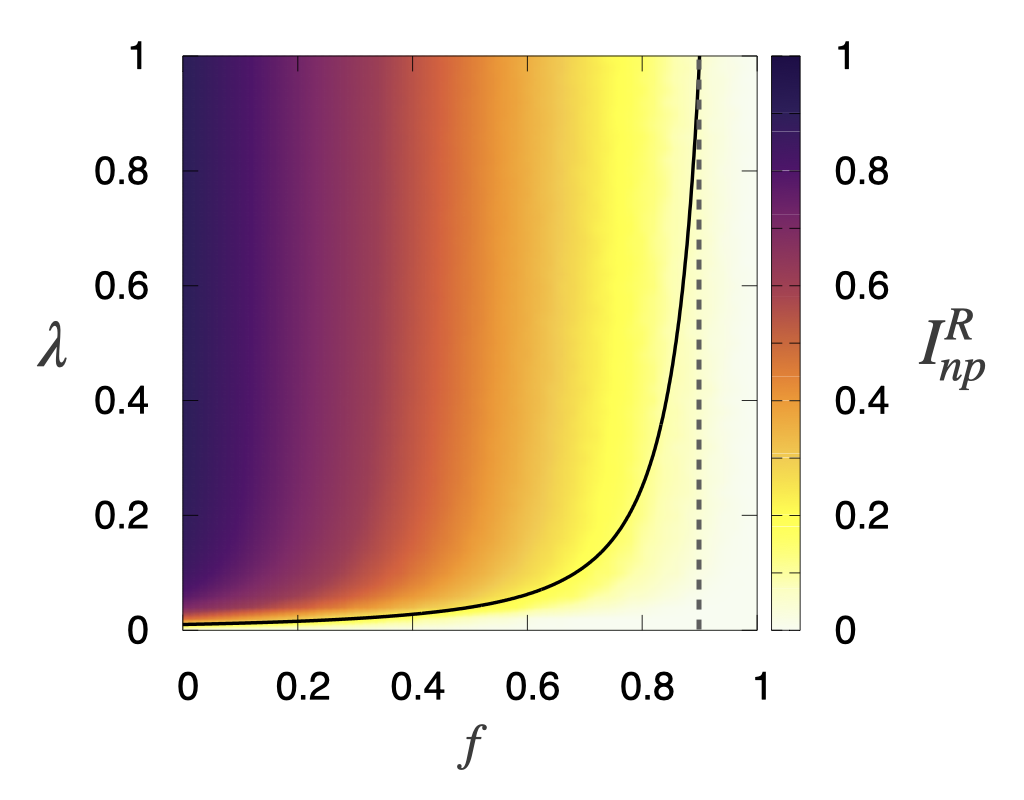}
	\caption{ Phase diagrams for the fraction of infected risky agents, $I^R_{np}$. The solid line indicates the analytical curve for the epidemic threshold, Eq.~(\ref{lanC}), while the dashed line indicates the critical value $f_c$, Eq.~(\ref{fc}). Parameters are set to $\delta=0$, $\gamma=0$, $\mu=0.1$, $c^R=1$ and $T^C=500$ on the ER network with $N=2000$ and $\langle k \rangle=10$.}
	\label{stub}
\end{figure}

We are particularly interested in the case that the two groups are complete zealots and thus the awareness gap is the largest possible one. With this aim, we assume the most effective protection, $\gamma=0$. In this special case, all concerned agents are protected and healthy ($P^C=1$ and $I^C=0$) and the disease spreads only among risky agents for which $P^R=0$. The question thus is to what extent the protection provided by concerned agents extends to the risky population. This question is answered by the phase diagram for the fraction of risky agents that are infected ($I^R_{np}$) as a function of $f$ and $\lambda$, shown in Fig.~\ref{stub}. Apart from the usual epidemic threshold $\lambda_c$ for small values of $f$ the diagram shows a clear transition from healthy to endemic phase when $f$ is varied. Again, we find a threshold value $f_c$ so that for $f>f_c$ risky individuals remain healthy for any value of $\lambda$ (although not taking any protection) in a similar way to the results shown in Figs.~\ref{resultsGams} and \ref{thre} when, for $f>f_c$, risky agents refrain from protecting themselves while being mostly susceptible sheltered by a kind of herd immunity provided by the protection taken by the concerned agents .

Let us now derive the expression of $f_c$ in the case of a system composed of concerned and risky zealots. Since risky agents never change their NP strategy, we conclude $\Gamma_{np\to p}^{R} =0$ and $\Gamma_{p\to np}^{R}=1$. Inserting these conditions in Eq.~(\ref{Inp}), we obtain:
\begin{equation}
	I^{i,R}_{np}(t+1) =  \left(1-I^{i,R}_{np}(t)\right)q^{i,R}_{np}(t) + (1-\mu)I^{i,R}_{np}(t)\;,
	\label{Stub1}
\end{equation}
where, according to Eq.~(\ref{qnp}) the infection probability for risky agents is given by:
\begin{equation}
	q^{i,R}_{np}(t) = 1 -\displaystyle\prod_{j=1}^{N} \big{[} 1 - \lambda A_{ij}^{RR} I^{j,\alpha}_{np} (t) \big{]}\;.
\end{equation}

In order to find the herd immunity threshold $f_c$, let us assume that our system has reached an equilibrium state and that, in this equilibrium, the probability that a risky (not protected) agent is infected is small: $I^{i,R}_{np}(t+1) =I^{i,R}_{np}(t) \equiv \epsilon^{R}_i$. In this limit we can write Eq.~(\ref{Stub1}) as:
\begin{equation}
	\epsilon^{R}_{i} =  \lambda\left(1-\epsilon^{R}_{i}\right)\sum_{j=1}^{N}A^{RR}_{ij}\epsilon^{R}_{j} + (1-\mu)\epsilon^{R}_{i}
	\label{Stub2}
\end{equation}
To satisfy this equation the following equality must hold:
\begin{equation}
	\frac{\lambda\Lambda_{max}({\bf A}^{RR})}{\mu}=1\;,
	\label{Stub3}
\end{equation}
where $\Lambda_{max}({\bf A}^{RR})$ is the maximum eigenvalue of matrix ${\bf A}^{RR}$. Now let us recall that ${\bf A}^{RR}$ is a $N\times N$ matrix that contains only those links connecting individuals of type $R$. Thus, considering that if two nodes $i$ and $j$ are connected then $A_{ij}=1$ (where ${\bf A}$ is the adjacency matrix of the network) and that the probability that these agents are both of type $R$ is $(1-f)^2$, we can approximate the elements of matrix ${\bf A}^{RR}$ as $A^{RR}_{ij}=(1-f)^2 A_{ij}$, so that ${\bf A}^{RR}$ can be read as the probability that given a network a link connected two $R$ agents. Thus, $\Lambda_{max}({\bf A^{RR}})=(1-f)^2\Lambda_{max}({\bf A})$ and Eq.~(\ref{Stub3}) can be written as:
\begin{equation}
	\frac{\lambda(1-f)^2\Lambda_{max}({\bf A})}{\mu}=1\;.
	\label{Stub4}
\end{equation}
Now, by noting that $\Lambda_{max}({\bf A})\simeq \langle k^2\rangle/\langle k\rangle$ \cite{SPECTRA,Restrepo} and considering that, for the Erd\"os-R\'enyi networks we can approximate  $ \langle k^2\rangle\sim\langle k\rangle^2$ we obtain the following expression for the epidemic threshold:
\begin{equation}
	\frac{\lambda(1-f)^2\langle k\rangle}{\mu}=1.
	\label{lanC}
\end{equation}
The former equation defines a curve $\lambda_c(f)$ (solid curve in Fig.~(\ref{stub}) that pinpoints the border between the disease-free regime and the epidemic one. Finally, we consider the maximum value of the epidemic threshold $\lambda_c=1$, and find the critical fraction $f_c$ of concerned population that gives herd-immunity to the risky one as:
\begin{equation}
	\frac{(1-f_c)^2\langle k\rangle}{\mu}= 1 \Rightarrow f_c=1-\sqrt{\frac{\mu}{\langle k\rangle}}\;.
	\label{fc}
\end{equation}
Given the sharp behavior of the curve $\lambda_c(f)$ near $f_c$ a coarse-grained description of the phase diagram is as follows. For $f < f_c$, the epidemic threshold of $\lambda_c$ is a function of the recovery probability and the fraction of the concerned agents. However, at critical point $f_c$, the epidemic threshold jumps to $1$. In other words, if the fraction of concerned people in the society exceeds a certain value, the disease-free phase is reachable for risky agents regardless of how large the probability of contagion is. The analytical curves are compatible with the numerical results obtained in the previous section highlighting the generality of the herd-immunity threshold $f_c$.

\section{Conclusions}
\label{secconclu}
In this work, we have studied the emergence of behavioural responses under the threat of the spread of a disease, how the adoption of protective measures impacts the course of epidemics, and how the interplay of different risk perceptions to the same disease shapes the collective response of the population.  In this framework, people decide to protect or not depending on the perceived risk of infection and the costs associated to the protection measures and the disease. The main novelty of this work is the partition of the population into two groups with different social attitudes: concerned (C) and risky (R). In particular, both the cost of taking the protective measures, $c$, and that of contracting the disease, $T$, are assumed to be different for concerned and risky agents so that $c^{C}< c^{R}$ and $T^{C}> T^{R}$.

Within this framework, and considering a random distribution of concerned and risky agents in homogeneous Erd\"os-R\'enyi graphs, we have focused on the identification of the different equilibria that appear as a result of the coexistence and interplay of the two populations. We have shown that when the population of concerned and risky players are identical, risky agents can capitalize on the protective effort of concerned ones, i.e. remain susceptible without being protected, provided the cost associated is less than a threshold or when the awareness gap is large enough. When the populations are not of equal size a herd-immunity effect over risky agents can be attained provided the fraction of concerned agents becomes larger than a threshold value. Other features such as the dynamical characterization or the coexistence equilibria or the effects of different mixing patterns  between concerned and risky populations will be addressed elsewhere \cite{Benja2}.

Our results, within the previous limitations, shed light on the difficulty of countering the existence of denialist minorities since they may find a false sense of security in the herd immunity created by the concerned majority. The danger of this effect is the potential spread of risky behavior. This propagation has not been considered here, since the populations of the two groups were considered static, and is left for the future work.

\section{Acknowledgements} {MK acknowledges the Ministry of Science, Research, and Technology of Iran. A.A. acknowledges the Spanish MINECO (Grant No. PGC2018-094754-B-C2), the Generalitat de Catalunya (Grants No. 2017SGR-896 and 2020PANDE00098), the Universitat Rovira i Virgili (Grant No. 2017PFR- URV-B2-41), and the ICREA Academia and the James S. Mc-Donnell Foundation (Grant No. 220020325). JGG acknowledges the Spanish MINECO (Grant No. FIS2017-87519-P), the Departamento de Industria e Innovaci\'on del Gobierno de Arag\'on and Fondo Social Europeo through (Grant No. E36-17R FENOL), and Fundaci\'on Ibercaja and Universidad de Zaragoza (Grant No. 224220). We acknowledge useful discussions with A. Reyna-Lara, D. Soriano-Pa\~nos and B. Steinegger.}

\bibliography{mybibfile}

\begin{thebibliography}{43}%
\makeatletter
\providecommand \@ifxundefined [1]{%
 \@ifx{#1\undefined}
}%
\providecommand \@ifnum [1]{%
 \ifnum #1\expandafter \@firstoftwo
 \else \expandafter \@secondoftwo
 \fi
}%
\providecommand \@ifx [1]{%
 \ifx #1\expandafter \@firstoftwo
 \else \expandafter \@secondoftwo
 \fi
}%
\providecommand \natexlab [1]{#1}%
\providecommand \enquote  [1]{``#1''}%
\providecommand \bibnamefont  [1]{#1}%
\providecommand \bibfnamefont [1]{#1}%
\providecommand \citenamefont [1]{#1}%
\providecommand \href@noop [0]{\@secondoftwo}%
\providecommand \href [0]{\begingroup \@sanitize@url \@href}%
\providecommand \@href[1]{\@@startlink{#1}\@@href}%
\providecommand \@@href[1]{\endgroup#1\@@endlink}%
\providecommand \@sanitize@url [0]{\catcode `\\12\catcode `\$12\catcode
  `\&12\catcode `\#12\catcode `\^12\catcode `\_12\catcode `\%12\relax}%
\providecommand \@@startlink[1]{}%
\providecommand \@@endlink[0]{}%
\providecommand \url  [0]{\begingroup\@sanitize@url \@url }%
\providecommand \@url [1]{\endgroup\@href {#1}{\urlprefix }}%
\providecommand \urlprefix  [0]{URL }%
\providecommand \Eprint [0]{\href }%
\providecommand \doibase [0]{http://dx.doi.org/}%
\providecommand \selectlanguage [0]{\@gobble}%
\providecommand \bibinfo  [0]{\@secondoftwo}%
\providecommand \bibfield  [0]{\@secondoftwo}%
\providecommand \translation [1]{[#1]}%
\providecommand \BibitemOpen [0]{}%
\providecommand \bibitemStop [0]{}%
\providecommand \bibitemNoStop [0]{.\EOS\space}%
\providecommand \EOS [0]{\spacefactor3000\relax}%
\providecommand \BibitemShut  [1]{\csname bibitem#1\endcsname}%
\let\auto@bib@innerbib\@empty
\bibitem [{\citenamefont {Anderson}\ and\ \citenamefont
  {May}(1992)}]{Anderson}%
  \BibitemOpen
  \bibfield  {author} {\bibinfo {author} {\bibfnamefont {R.}~\bibnamefont
  {Anderson}}\ and\ \bibinfo {author} {\bibfnamefont {R.}~\bibnamefont {May}},\
  }\href@noop {} {\emph {\bibinfo {title} {Infectious Diseases of Humans.
  Dynamics and Control}}}\ (\bibinfo  {publisher} {Oxford University Press},\
  \bibinfo {year} {1992})\BibitemShut {NoStop}%
\bibitem [{\citenamefont {Hethcote}(2000)}]{Hethcote}%
  \BibitemOpen
  \bibfield  {author} {\bibinfo {author} {\bibfnamefont {H.~W.}\ \bibnamefont
  {Hethcote}},\ }\href@noop {} {\bibfield  {journal} {\bibinfo  {journal} {SIAM
  Rev.}\ }\textbf {\bibinfo {volume} {42}},\ \bibinfo {pages} {599} (\bibinfo
  {year} {2000})}\BibitemShut {NoStop}%
\bibitem [{\citenamefont {Keeling}\ and\ \citenamefont
  {Rohani}(2007)}]{Keeling}%
  \BibitemOpen
  \bibfield  {author} {\bibinfo {author} {\bibfnamefont {M.}~\bibnamefont
  {Keeling}}\ and\ \bibinfo {author} {\bibfnamefont {P.}~\bibnamefont
  {Rohani}},\ }\href@noop {} {\emph {\bibinfo {title} {Modeling Infectious
  Diseases in Humans and Animals}}}\ (\bibinfo  {publisher} {Princeton
  University Press},\ \bibinfo {year} {2007})\BibitemShut {NoStop}%
\bibitem [{\citenamefont {Newman}(2002)}]{Newman}%
  \BibitemOpen
  \bibfield  {author} {\bibinfo {author} {\bibfnamefont {M.~E.~J.}\
  \bibnamefont {Newman}},\ }\href {\doibase 10.1103/PhysRevE.66.016128}
  {\bibfield  {journal} {\bibinfo  {journal} {Phys. Rev. E}\ }\textbf {\bibinfo
  {volume} {66}},\ \bibinfo {pages} {016128} (\bibinfo {year}
  {2002})}\BibitemShut {NoStop}%
\bibitem [{\citenamefont {Schneider}\ \emph {et~al.}(2011)\citenamefont
  {Schneider}, \citenamefont {Mihaljev}, \citenamefont {Havlin},\ and\
  \citenamefont {Herrmann}}]{Target}%
  \BibitemOpen
  \bibfield  {author} {\bibinfo {author} {\bibfnamefont {C.~M.}\ \bibnamefont
  {Schneider}}, \bibinfo {author} {\bibfnamefont {T.}~\bibnamefont {Mihaljev}},
  \bibinfo {author} {\bibfnamefont {S.}~\bibnamefont {Havlin}}, \ and\ \bibinfo
  {author} {\bibfnamefont {H.~J.}\ \bibnamefont {Herrmann}},\ }\href {\doibase
  10.1103/PhysRevE.84.061911} {\bibfield  {journal} {\bibinfo  {journal} {Phys.
  Rev. E}\ }\textbf {\bibinfo {volume} {84}},\ \bibinfo {pages} {061911}
  (\bibinfo {year} {2011})}\BibitemShut {NoStop}%
\bibitem [{\citenamefont {Wang}\ \emph {et~al.}(2017)\citenamefont {Wang},
  \citenamefont {Moreno}, \citenamefont {Boccaletti},\ and\ \citenamefont
  {Perc}}]{Wang}%
  \BibitemOpen
  \bibfield  {author} {\bibinfo {author} {\bibfnamefont {Z.}~\bibnamefont
  {Wang}}, \bibinfo {author} {\bibfnamefont {Y.}~\bibnamefont {Moreno}},
  \bibinfo {author} {\bibfnamefont {S.}~\bibnamefont {Boccaletti}}, \ and\
  \bibinfo {author} {\bibfnamefont {M.}~\bibnamefont {Perc}},\ }\href {\doibase
  https://doi.org/10.1016/j.chaos.2017.06.004} {\bibfield  {journal} {\bibinfo
  {journal} {Chaos, Solitons \& Fractals}\ }\textbf {\bibinfo {volume} {103}},\
  \bibinfo {pages} {177 } (\bibinfo {year} {2017})}\BibitemShut {NoStop}%
\bibitem [{\citenamefont {Cohen}\ \emph {et~al.}(2003)\citenamefont {Cohen},
  \citenamefont {Havlin},\ and\ \citenamefont {ben Avraham}}]{Acquaintance}%
  \BibitemOpen
  \bibfield  {author} {\bibinfo {author} {\bibfnamefont {R.}~\bibnamefont
  {Cohen}}, \bibinfo {author} {\bibfnamefont {S.}~\bibnamefont {Havlin}}, \
  and\ \bibinfo {author} {\bibfnamefont {D.}~\bibnamefont {ben Avraham}},\
  }\href {\doibase 10.1103/PhysRevLett.91.247901} {\bibfield  {journal}
  {\bibinfo  {journal} {Phys. Rev. Lett.}\ }\textbf {\bibinfo {volume} {91}},\
  \bibinfo {pages} {247901} (\bibinfo {year} {2003})}\BibitemShut {NoStop}%
\bibitem [{\citenamefont {Ye}\ \emph {et~al.}(2020)\citenamefont {Ye},
  \citenamefont {Zhang}, \citenamefont {Cao},\ and\ \citenamefont
  {Zeng}}]{Qingpeng}%
  \BibitemOpen
  \bibfield  {author} {\bibinfo {author} {\bibfnamefont {Y.}~\bibnamefont
  {Ye}}, \bibinfo {author} {\bibfnamefont {Q.}~\bibnamefont {Zhang}}, \bibinfo
  {author} {\bibfnamefont {Z.}~\bibnamefont {Cao}}, \ and\ \bibinfo {author}
  {\bibfnamefont {D.~D.}\ \bibnamefont {Zeng}},\ }\href@noop {} {\bibfield
  {journal} {\bibinfo  {journal} {arXiv:2011.14255}\ } (\bibinfo {year}
  {2020})}\BibitemShut {NoStop}%
\bibitem [{\citenamefont {Gandon}\ \emph {et~al.}(2001)\citenamefont {Gandon},
  \citenamefont {Mackinnon}, \citenamefont {Nee},\ and\ \citenamefont
  {Read}}]{Gardon}%
  \BibitemOpen
  \bibfield  {author} {\bibinfo {author} {\bibfnamefont {S.}~\bibnamefont
  {Gandon}}, \bibinfo {author} {\bibfnamefont {M.~J.}\ \bibnamefont
  {Mackinnon}}, \bibinfo {author} {\bibfnamefont {S.}~\bibnamefont {Nee}}, \
  and\ \bibinfo {author} {\bibfnamefont {A.~F.}\ \bibnamefont {Read}},\ }\href
  {\doibase 10.1038/414751a} {\bibfield  {journal} {\bibinfo  {journal}
  {Nature}\ }\textbf {\bibinfo {volume} {414}},\ \bibinfo {pages} {751}
  (\bibinfo {year} {2001})}\BibitemShut {NoStop}%
\bibitem [{\citenamefont {Kribs-Zaleta}\ and\ \citenamefont
  {Velasco-Hernández}(2000)}]{Zaleta}%
  \BibitemOpen
  \bibfield  {author} {\bibinfo {author} {\bibfnamefont {C.~M.}\ \bibnamefont
  {Kribs-Zaleta}}\ and\ \bibinfo {author} {\bibfnamefont {J.~X.}\ \bibnamefont
  {Velasco-Hernández}},\ }\href {\doibase
  https://doi.org/10.1016/S0025-5564(00)00003-1} {\bibfield  {journal}
  {\bibinfo  {journal} {Mathematical Biosciences}\ }\textbf {\bibinfo {volume}
  {164}},\ \bibinfo {pages} {183 } (\bibinfo {year} {2000})}\BibitemShut
  {NoStop}%
\bibitem [{\citenamefont {Peng}\ \emph {et~al.}(2013)\citenamefont {Peng},
  \citenamefont {Xu}, \citenamefont {Fu},\ and\ \citenamefont {Zhou}}]{Peng}%
  \BibitemOpen
  \bibfield  {author} {\bibinfo {author} {\bibfnamefont {X.-L.}\ \bibnamefont
  {Peng}}, \bibinfo {author} {\bibfnamefont {X.-J.}\ \bibnamefont {Xu}},
  \bibinfo {author} {\bibfnamefont {X.}~\bibnamefont {Fu}}, \ and\ \bibinfo
  {author} {\bibfnamefont {T.}~\bibnamefont {Zhou}},\ }\href {\doibase
  10.1103/PhysRevE.87.022813} {\bibfield  {journal} {\bibinfo  {journal} {Phys.
  Rev. E}\ }\textbf {\bibinfo {volume} {87}},\ \bibinfo {pages} {022813}
  (\bibinfo {year} {2013})}\BibitemShut {NoStop}%
\bibitem [{\citenamefont {Peng}\ \emph {et~al.}(2016)\citenamefont {Peng},
  \citenamefont {Xu}, \citenamefont {Small}, \citenamefont {Fu},\ and\
  \citenamefont {Jin}}]{Long}%
  \BibitemOpen
  \bibfield  {author} {\bibinfo {author} {\bibfnamefont {X.-L.}\ \bibnamefont
  {Peng}}, \bibinfo {author} {\bibfnamefont {X.-J.}\ \bibnamefont {Xu}},
  \bibinfo {author} {\bibfnamefont {M.}~\bibnamefont {Small}}, \bibinfo
  {author} {\bibfnamefont {X.}~\bibnamefont {Fu}}, \ and\ \bibinfo {author}
  {\bibfnamefont {Z.}~\bibnamefont {Jin}},\ }\href {\doibase
  10.1007/s00285-016-1007-3} {\bibfield  {journal} {\bibinfo  {journal}
  {Journal of Mathematical Biology}\ }\textbf {\bibinfo {volume} {73}},\
  \bibinfo {pages} {1561} (\bibinfo {year} {2016})}\BibitemShut {NoStop}%
\bibitem [{\citenamefont {Chen}\ and\ \citenamefont {Fu}(2019)}]{Chen}%
  \BibitemOpen
  \bibfield  {author} {\bibinfo {author} {\bibfnamefont {X.}~\bibnamefont
  {Chen}}\ and\ \bibinfo {author} {\bibfnamefont {F.}~\bibnamefont {Fu}},\
  }\href {\doibase 10.1098/rspb.2018.2406} {\bibfield  {journal} {\bibinfo
  {journal} {Proceedings of the Royal Society B: Biological Sciences}\ }\textbf
  {\bibinfo {volume} {286}},\ \bibinfo {pages} {20182406} (\bibinfo {year}
  {2019})}\BibitemShut {NoStop}%
\bibitem [{\citenamefont {Khanjanianpak}\ \emph {et~al.}(2020)\citenamefont
  {Khanjanianpak}, \citenamefont {Azimi-Tafreshi},\ and\ \citenamefont
  {Castellano}}]{Mozhgan}%
  \BibitemOpen
  \bibfield  {author} {\bibinfo {author} {\bibfnamefont {M.}~\bibnamefont
  {Khanjanianpak}}, \bibinfo {author} {\bibfnamefont {N.}~\bibnamefont
  {Azimi-Tafreshi}}, \ and\ \bibinfo {author} {\bibfnamefont {C.}~\bibnamefont
  {Castellano}},\ }\href@noop {} {\bibfield  {journal} {\bibinfo  {journal}
  {Phys. Rev. E}\ }\textbf {\bibinfo {volume} {101}},\ \bibinfo {pages}
  {062306} (\bibinfo {year} {2020})}\BibitemShut {NoStop}%
\bibitem [{\citenamefont {Kato}\ \emph {et~al.}(2011)\citenamefont {Kato},
  \citenamefont {Tainaka}, \citenamefont {Sone}, \citenamefont {Morita},
  \citenamefont {Iida},\ and\ \citenamefont {Yoshimura}}]{Fuminori}%
  \BibitemOpen
  \bibfield  {author} {\bibinfo {author} {\bibfnamefont {F.}~\bibnamefont
  {Kato}}, \bibinfo {author} {\bibfnamefont {K.-i.}\ \bibnamefont {Tainaka}},
  \bibinfo {author} {\bibfnamefont {S.}~\bibnamefont {Sone}}, \bibinfo {author}
  {\bibfnamefont {S.}~\bibnamefont {Morita}}, \bibinfo {author} {\bibfnamefont
  {H.}~\bibnamefont {Iida}}, \ and\ \bibinfo {author} {\bibfnamefont
  {J.}~\bibnamefont {Yoshimura}},\ }\href@noop {} {\bibfield  {journal}
  {\bibinfo  {journal} {Sci. Rep.}\ }\textbf {\bibinfo {volume} {1}},\ \bibinfo
  {pages} {1} (\bibinfo {year} {2011})}\BibitemShut {NoStop}%
\bibitem [{\citenamefont {Gosak}\ \emph
  {et~al.}(2021{\natexlab{a}})\citenamefont {Gosak}, \citenamefont {Kraemer},
  \citenamefont {Nax}, \citenamefont {Perc},\ and\ \citenamefont
  {Pradelski}}]{gosak1}%
  \BibitemOpen
  \bibfield  {author} {\bibinfo {author} {\bibfnamefont {M.}~\bibnamefont
  {Gosak}}, \bibinfo {author} {\bibfnamefont {M.~U.}\ \bibnamefont {Kraemer}},
  \bibinfo {author} {\bibfnamefont {H.~H.}\ \bibnamefont {Nax}}, \bibinfo
  {author} {\bibfnamefont {M.}~\bibnamefont {Perc}}, \ and\ \bibinfo {author}
  {\bibfnamefont {B.~S.}\ \bibnamefont {Pradelski}},\ }\href@noop {} {\bibfield
   {journal} {\bibinfo  {journal} {Scientific reports}\ }\textbf {\bibinfo
  {volume} {11}},\ \bibinfo {pages} {1} (\bibinfo {year}
  {2021}{\natexlab{a}})}\BibitemShut {NoStop}%
\bibitem [{\citenamefont {Gosak}\ \emph
  {et~al.}(2021{\natexlab{b}})\citenamefont {Gosak}, \citenamefont {Duh},
  \citenamefont {Markovi{\v{c}}},\ and\ \citenamefont {Perc}}]{gosak2}%
  \BibitemOpen
  \bibfield  {author} {\bibinfo {author} {\bibfnamefont {M.}~\bibnamefont
  {Gosak}}, \bibinfo {author} {\bibfnamefont {M.}~\bibnamefont {Duh}}, \bibinfo
  {author} {\bibfnamefont {R.}~\bibnamefont {Markovi{\v{c}}}}, \ and\ \bibinfo
  {author} {\bibfnamefont {M.}~\bibnamefont {Perc}},\ }\href@noop {} {\bibfield
   {journal} {\bibinfo  {journal} {New Journal of Physics}\ }\textbf {\bibinfo
  {volume} {23}},\ \bibinfo {pages} {043039} (\bibinfo {year}
  {2021}{\natexlab{b}})}\BibitemShut {NoStop}%
\bibitem [{\citenamefont {Verelst}\ \emph {et~al.}(2016)\citenamefont
  {Verelst}, \citenamefont {Willem},\ and\ \citenamefont {Beutels}}]{Fredrik}%
  \BibitemOpen
  \bibfield  {author} {\bibinfo {author} {\bibfnamefont {F.}~\bibnamefont
  {Verelst}}, \bibinfo {author} {\bibfnamefont {L.}~\bibnamefont {Willem}}, \
  and\ \bibinfo {author} {\bibfnamefont {P.}~\bibnamefont {Beutels}},\
  }\href@noop {} {\bibfield  {journal} {\bibinfo  {journal} {J. R. Soc.
  Interface}\ }\textbf {\bibinfo {volume} {13}},\ \bibinfo {pages} {20160820}
  (\bibinfo {year} {2016})}\BibitemShut {NoStop}%
\bibitem [{\citenamefont {Bauch}\ \emph {et~al.}(2003)\citenamefont {Bauch},
  \citenamefont {Galvani},\ and\ \citenamefont {Earn}}]{Bauch1}%
  \BibitemOpen
  \bibfield  {author} {\bibinfo {author} {\bibfnamefont {C.}~\bibnamefont
  {Bauch}}, \bibinfo {author} {\bibfnamefont {A.}~\bibnamefont {Galvani}}, \
  and\ \bibinfo {author} {\bibfnamefont {D.}~\bibnamefont {Earn}},\ }\href@noop
  {} {\bibfield  {journal} {\bibinfo  {journal} {Proc. Nat. Acad. Sci. USA}\
  }\textbf {\bibinfo {volume} {100}},\ \bibinfo {pages} {10564} (\bibinfo
  {year} {2003})}\BibitemShut {NoStop}%
\bibitem [{\citenamefont {Bauch}\ and\ \citenamefont {Earn}(2004)}]{Bauch2}%
  \BibitemOpen
  \bibfield  {author} {\bibinfo {author} {\bibfnamefont {C.}~\bibnamefont
  {Bauch}}\ and\ \bibinfo {author} {\bibfnamefont {D.}~\bibnamefont {Earn}},\
  }\href@noop {} {\bibfield  {journal} {\bibinfo  {journal} {Proc. Nat. Acad.
  Sci. USA}\ }\textbf {\bibinfo {volume} {101}},\ \bibinfo {pages} {13391}
  (\bibinfo {year} {2004})}\BibitemShut {NoStop}%
\bibitem [{\citenamefont {Wang}\ \emph {et~al.}(2016)\citenamefont {Wang},
  \citenamefont {Bauch}, \citenamefont {Bhattacharyya}, \citenamefont
  {d'Onofrio}, \citenamefont {Manfredi}, \citenamefont {Perc}, \citenamefont
  {Perra}, \citenamefont {Salath{\'e}},\ and\ \citenamefont {Zhao}}]{Zhen}%
  \BibitemOpen
  \bibfield  {author} {\bibinfo {author} {\bibfnamefont {Z.}~\bibnamefont
  {Wang}}, \bibinfo {author} {\bibfnamefont {C.~T.}\ \bibnamefont {Bauch}},
  \bibinfo {author} {\bibfnamefont {S.}~\bibnamefont {Bhattacharyya}}, \bibinfo
  {author} {\bibfnamefont {A.}~\bibnamefont {d'Onofrio}}, \bibinfo {author}
  {\bibfnamefont {P.}~\bibnamefont {Manfredi}}, \bibinfo {author}
  {\bibfnamefont {M.}~\bibnamefont {Perc}}, \bibinfo {author} {\bibfnamefont
  {N.}~\bibnamefont {Perra}}, \bibinfo {author} {\bibfnamefont
  {M.}~\bibnamefont {Salath{\'e}}}, \ and\ \bibinfo {author} {\bibfnamefont
  {D.}~\bibnamefont {Zhao}},\ }\href@noop {} {\bibfield  {journal} {\bibinfo
  {journal} {Phys. Rep.}\ }\textbf {\bibinfo {volume} {664}},\ \bibinfo {pages}
  {1} (\bibinfo {year} {2016})}\BibitemShut {NoStop}%
\bibitem [{\citenamefont {Chang}\ \emph {et~al.}(2020)\citenamefont {Chang},
  \citenamefont {Piraveenan}, \citenamefont {Pattison},\ and\ \citenamefont
  {Prokopenko}}]{Sheryl}%
  \BibitemOpen
  \bibfield  {author} {\bibinfo {author} {\bibfnamefont {S.~L.}\ \bibnamefont
  {Chang}}, \bibinfo {author} {\bibfnamefont {M.}~\bibnamefont {Piraveenan}},
  \bibinfo {author} {\bibfnamefont {P.}~\bibnamefont {Pattison}}, \ and\
  \bibinfo {author} {\bibfnamefont {M.}~\bibnamefont {Prokopenko}},\
  }\href@noop {} {\bibfield  {journal} {\bibinfo  {journal} {J. Biol. Dyn.}\
  }\textbf {\bibinfo {volume} {14}},\ \bibinfo {pages} {57} (\bibinfo {year}
  {2020})}\BibitemShut {NoStop}%
\bibitem [{\citenamefont {Tanimoto}(2019)}]{Tanimoto}%
  \BibitemOpen
  \bibfield  {author} {\bibinfo {author} {\bibfnamefont {J.}~\bibnamefont
  {Tanimoto}},\ }\href@noop {} {\bibfield  {journal} {\bibinfo  {journal}
  {\textit{Evolutionary Economics: Analysis of Traffic Flow and Epidemics}}\ }
  (\bibinfo {year} {2019})}\BibitemShut {NoStop}%
\bibitem [{\citenamefont {Ndeffo~Mbah}\ \emph {et~al.}(2012)\citenamefont
  {Ndeffo~Mbah}, \citenamefont {Liu}, \citenamefont {Bauch}, \citenamefont
  {Tekel}, \citenamefont {Medlock}, \citenamefont {Meyers},\ and\ \citenamefont
  {Galvani}}]{Martial}%
  \BibitemOpen
  \bibfield  {author} {\bibinfo {author} {\bibfnamefont {M.~L.}\ \bibnamefont
  {Ndeffo~Mbah}}, \bibinfo {author} {\bibfnamefont {J.}~\bibnamefont {Liu}},
  \bibinfo {author} {\bibfnamefont {C.~T.}\ \bibnamefont {Bauch}}, \bibinfo
  {author} {\bibfnamefont {Y.~I.}\ \bibnamefont {Tekel}}, \bibinfo {author}
  {\bibfnamefont {J.}~\bibnamefont {Medlock}}, \bibinfo {author} {\bibfnamefont
  {L.~A.}\ \bibnamefont {Meyers}}, \ and\ \bibinfo {author} {\bibfnamefont
  {A.~P.}\ \bibnamefont {Galvani}},\ }\href@noop {} {\bibfield  {journal}
  {\bibinfo  {journal} {PLoS Comput. Biol.}\ }\textbf {\bibinfo {volume} {8}},\
  \bibinfo {pages} {e1002469} (\bibinfo {year} {2012})}\BibitemShut {NoStop}%
\bibitem [{\citenamefont {Brevan}\ \emph {et~al.}(2007)\citenamefont {Brevan},
  \citenamefont {Vardavas},\ and\ \citenamefont {Blower}}]{Brevan}%
  \BibitemOpen
  \bibfield  {author} {\bibinfo {author} {\bibfnamefont {R.}~\bibnamefont
  {Brevan}}, \bibinfo {author} {\bibfnamefont {R.}~\bibnamefont {Vardavas}}, \
  and\ \bibinfo {author} {\bibfnamefont {S.}~\bibnamefont {Blower}},\
  }\href@noop {} {\bibfield  {journal} {\bibinfo  {journal} {Phys. Rev. E}\
  }\textbf {\bibinfo {volume} {76}},\ \bibinfo {pages} {031127} (\bibinfo
  {year} {2007})}\BibitemShut {NoStop}%
\bibitem [{\citenamefont {Zhang}\ \emph {et~al.}(2010)\citenamefont {Zhang},
  \citenamefont {Zhang}, \citenamefont {Zhou}, \citenamefont {Small},\ and\
  \citenamefont {Wang}}]{Zhang}%
  \BibitemOpen
  \bibfield  {author} {\bibinfo {author} {\bibfnamefont {H.}~\bibnamefont
  {Zhang}}, \bibinfo {author} {\bibfnamefont {J.}~\bibnamefont {Zhang}},
  \bibinfo {author} {\bibfnamefont {C.}~\bibnamefont {Zhou}}, \bibinfo {author}
  {\bibfnamefont {M.}~\bibnamefont {Small}}, \ and\ \bibinfo {author}
  {\bibfnamefont {B.}~\bibnamefont {Wang}},\ }\href@noop {} {\bibfield
  {journal} {\bibinfo  {journal} {New J. Phys.}\ }\textbf {\bibinfo {volume}
  {12}},\ \bibinfo {pages} {023015} (\bibinfo {year} {2010})}\BibitemShut
  {NoStop}%
\bibitem [{\citenamefont {Perisic}\ and\ \citenamefont
  {Bauch}(2008)}]{Perisic}%
  \BibitemOpen
  \bibfield  {author} {\bibinfo {author} {\bibfnamefont {A.}~\bibnamefont
  {Perisic}}\ and\ \bibinfo {author} {\bibfnamefont {C.}~\bibnamefont
  {Bauch}},\ }\href@noop {} {\bibfield  {journal} {\bibinfo  {journal} {PLoS
  Comput. Biol.}\ }\textbf {\bibinfo {volume} {5}},\ \bibinfo {pages}
  {e1000280} (\bibinfo {year} {2008})}\BibitemShut {NoStop}%
\bibitem [{\citenamefont {Reluga}(2010)}]{Timothy}%
  \BibitemOpen
  \bibfield  {author} {\bibinfo {author} {\bibfnamefont {T.~C.}\ \bibnamefont
  {Reluga}},\ }\href@noop {} {\bibfield  {journal} {\bibinfo  {journal} {PLoS
  Comput. Biol.}\ }\textbf {\bibinfo {volume} {6}},\ \bibinfo {pages}
  {e1000793} (\bibinfo {year} {2010})}\BibitemShut {NoStop}%
\bibitem [{\citenamefont {Wells}\ \emph {et~al.}(2013)\citenamefont {Wells},
  \citenamefont {Klein},\ and\ \citenamefont {Bauch}}]{Wells}%
  \BibitemOpen
  \bibfield  {author} {\bibinfo {author} {\bibfnamefont {C.}~\bibnamefont
  {Wells}}, \bibinfo {author} {\bibfnamefont {E.}~\bibnamefont {Klein}}, \ and\
  \bibinfo {author} {\bibfnamefont {C.}~\bibnamefont {Bauch}},\ }\href@noop {}
  {\bibfield  {journal} {\bibinfo  {journal} {PLoS Comput. Biol.}\ }\textbf
  {\bibinfo {volume} {9}},\ \bibinfo {pages} {e1002945} (\bibinfo {year}
  {2013})}\BibitemShut {NoStop}%
\bibitem [{\citenamefont {Cardillo}\ \emph {et~al.}(2013)\citenamefont
  {Cardillo}, \citenamefont {Reyes-Su{\'a}rez}, \citenamefont {Naranjo},\ and\
  \citenamefont {G{\'o}mez-Gardenes}}]{Alessio}%
  \BibitemOpen
  \bibfield  {author} {\bibinfo {author} {\bibfnamefont {A.}~\bibnamefont
  {Cardillo}}, \bibinfo {author} {\bibfnamefont {C.}~\bibnamefont
  {Reyes-Su{\'a}rez}}, \bibinfo {author} {\bibfnamefont {F.}~\bibnamefont
  {Naranjo}}, \ and\ \bibinfo {author} {\bibfnamefont {J.}~\bibnamefont
  {G{\'o}mez-Gardenes}},\ }\href@noop {} {\bibfield  {journal} {\bibinfo
  {journal} {Phys. Rev. E}\ }\textbf {\bibinfo {volume} {88}},\ \bibinfo
  {pages} {032803} (\bibinfo {year} {2013})}\BibitemShut {NoStop}%
\bibitem [{\citenamefont {Steinegger}\ \emph {et~al.}(2018)\citenamefont
  {Steinegger}, \citenamefont {Cardillo}, \citenamefont {Rios}, \citenamefont
  {G\'omez-Garde\~nes},\ and\ \citenamefont {Arenas}}]{Steinegger}%
  \BibitemOpen
  \bibfield  {author} {\bibinfo {author} {\bibfnamefont {B.}~\bibnamefont
  {Steinegger}}, \bibinfo {author} {\bibfnamefont {A.}~\bibnamefont
  {Cardillo}}, \bibinfo {author} {\bibfnamefont {P.~D.~L.}\ \bibnamefont
  {Rios}}, \bibinfo {author} {\bibfnamefont {J.}~\bibnamefont
  {G\'omez-Garde\~nes}}, \ and\ \bibinfo {author} {\bibfnamefont
  {A.}~\bibnamefont {Arenas}},\ }\href {\doibase 10.1103/PhysRevE.97.032308}
  {\bibfield  {journal} {\bibinfo  {journal} {Phys. Rev. E}\ }\textbf {\bibinfo
  {volume} {97}},\ \bibinfo {pages} {032308} (\bibinfo {year}
  {2018})}\BibitemShut {NoStop}%
\bibitem [{\citenamefont {Amaral}\ \emph {et~al.}(2020)\citenamefont {Amaral},
  \citenamefont {de~Oliveira},\ and\ \citenamefont {Javarone}}]{Marco}%
  \BibitemOpen
  \bibfield  {author} {\bibinfo {author} {\bibfnamefont {M.}~\bibnamefont
  {Amaral}}, \bibinfo {author} {\bibfnamefont {M.}~\bibnamefont {de~Oliveira}},
  \ and\ \bibinfo {author} {\bibfnamefont {M.}~\bibnamefont {Javarone}},\
  }\href@noop {} {\bibfield  {journal} {\bibinfo  {journal} {Chaos, Solitons \&
  Fractals}\ }\textbf {\bibinfo {volume} {143}},\ \bibinfo {pages} {110616}
  (\bibinfo {year} {2020})}\BibitemShut {NoStop}%
\bibitem [{\citenamefont {Steinegger}\ \emph {et~al.}(2020)\citenamefont
  {Steinegger}, \citenamefont {Arenas}, \citenamefont
  {G{\'o}mez-Garde{\~n}es},\ and\ \citenamefont {Granell}}]{Benjamin}%
  \BibitemOpen
  \bibfield  {author} {\bibinfo {author} {\bibfnamefont {B.}~\bibnamefont
  {Steinegger}}, \bibinfo {author} {\bibfnamefont {A.}~\bibnamefont {Arenas}},
  \bibinfo {author} {\bibfnamefont {J.}~\bibnamefont {G{\'o}mez-Garde{\~n}es}},
  \ and\ \bibinfo {author} {\bibfnamefont {C.}~\bibnamefont {Granell}},\
  }\href@noop {} {\bibfield  {journal} {\bibinfo  {journal} {Phys. Rev.
  Research}\ }\textbf {\bibinfo {volume} {2}},\ \bibinfo {pages} {023181}
  (\bibinfo {year} {2020})}\BibitemShut {NoStop}%
\bibitem [{\citenamefont {Bavel}\ \emph {et~al.}(2020)\citenamefont {Bavel},
  \citenamefont {Baicker}, \citenamefont {Boggio},\ and\ \citenamefont {et.
  al}}]{Bavel}%
  \BibitemOpen
  \bibfield  {author} {\bibinfo {author} {\bibfnamefont {J.}~\bibnamefont
  {Bavel}}, \bibinfo {author} {\bibfnamefont {K.}~\bibnamefont {Baicker}},
  \bibinfo {author} {\bibfnamefont {P.}~\bibnamefont {Boggio}}, \ and\ \bibinfo
  {author} {\bibnamefont {et. al}},\ }\href {\doibase
  10.1038/s41562-020-0884-z} {\bibfield  {journal} {\bibinfo  {journal} {Nat.
  Hum. Behav.}\ }\textbf {\bibinfo {volume} {4}},\ \bibinfo {pages} {460}
  (\bibinfo {year} {2020})}\BibitemShut {NoStop}%
\bibitem [{\citenamefont {Tanaka}\ \emph {et~al.}(2002)\citenamefont {Tanaka},
  \citenamefont {Kumm},\ and\ \citenamefont {Feldman}}]{Mark}%
  \BibitemOpen
  \bibfield  {author} {\bibinfo {author} {\bibfnamefont {M.~M.}\ \bibnamefont
  {Tanaka}}, \bibinfo {author} {\bibfnamefont {J.}~\bibnamefont {Kumm}}, \ and\
  \bibinfo {author} {\bibfnamefont {M.~W.}\ \bibnamefont {Feldman}},\ }\href
  {\doibase https://doi.org/10.1006/tpbi.2002.1585} {\bibfield  {journal}
  {\bibinfo  {journal} {Theor Popul. Biol.}\ }\textbf {\bibinfo {volume}
  {62}},\ \bibinfo {pages} {111} (\bibinfo {year} {2002})}\BibitemShut
  {NoStop}%
\bibitem [{\citenamefont {Wang}\ and\ \citenamefont {Xia}(2020)}]{Zhishuang}%
  \BibitemOpen
  \bibfield  {author} {\bibinfo {author} {\bibfnamefont {Z.}~\bibnamefont
  {Wang}}\ and\ \bibinfo {author} {\bibfnamefont {C.}~\bibnamefont {Xia}},\
  }\href {\doibase https://doi.org/10.1007/s11071-020-06021-7} {\bibfield
  {journal} {\bibinfo  {journal} {Nonlinear Dyn.}\ }\textbf {\bibinfo {volume}
  {102}},\ \bibinfo {pages} {3039} (\bibinfo {year} {2020})}\BibitemShut
  {NoStop}%
\bibitem [{\citenamefont {G\'omez}\ \emph {et~al.}(2011)\citenamefont
  {G\'omez}, \citenamefont {G\'omez-Garde{\~{n}}es}, \citenamefont {Moreno},\
  and\ \citenamefont {Arenas}}]{GOMEZ}%
  \BibitemOpen
  \bibfield  {author} {\bibinfo {author} {\bibfnamefont {S.}~\bibnamefont
  {G\'omez}}, \bibinfo {author} {\bibfnamefont {J.}~\bibnamefont
  {G\'omez-Garde{\~{n}}es}}, \bibinfo {author} {\bibfnamefont {Y.}~\bibnamefont
  {Moreno}}, \ and\ \bibinfo {author} {\bibfnamefont {A.}~\bibnamefont
  {Arenas}},\ }\href@noop {} {\bibfield  {journal} {\bibinfo  {journal} {Phys.
  Rev. E}\ }\textbf {\bibinfo {volume} {84}},\ \bibinfo {pages} {036105}
  (\bibinfo {year} {2011})}\BibitemShut {NoStop}%
\bibitem [{\citenamefont {Hauert}\ and\ \citenamefont
  {Doebeli}(2004)}]{Hauert04}%
  \BibitemOpen
  \bibfield  {author} {\bibinfo {author} {\bibfnamefont {C.}~\bibnamefont
  {Hauert}}\ and\ \bibinfo {author} {\bibfnamefont {M.}~\bibnamefont
  {Doebeli}},\ }\href {\doibase https://doi.org/10.1038/nature02360} {\bibfield
   {journal} {\bibinfo  {journal} {Nature}\ }\textbf {\bibinfo {volume}
  {428}},\ \bibinfo {pages} {643} (\bibinfo {year} {2004})}\BibitemShut
  {NoStop}%
\bibitem [{\citenamefont {Perez-Roca}\ \emph {et~al.}(2009)\citenamefont
  {Perez-Roca}, \citenamefont {Cuesta},\ and\ \citenamefont
  {S\'anchez}}]{Roca}%
  \BibitemOpen
  \bibfield  {author} {\bibinfo {author} {\bibfnamefont {C.}~\bibnamefont
  {Perez-Roca}}, \bibinfo {author} {\bibfnamefont {J.}~\bibnamefont {Cuesta}},
  \ and\ \bibinfo {author} {\bibfnamefont {A.}~\bibnamefont {S\'anchez}},\
  }\href {\doibase https://doi.org/10.1007/s11071-020-06021-7} {\bibfield
  {journal} {\bibinfo  {journal} {Physics of Life Rev.}\ }\textbf {\bibinfo
  {volume} {6}},\ \bibinfo {pages} {208} (\bibinfo {year} {2009})}\BibitemShut
  {NoStop}%
\bibitem [{\citenamefont {Cardillo}\ \emph {et~al.}(2010)\citenamefont
  {Cardillo}, \citenamefont {G\'omez-Garde{\~n}es}, \citenamefont {Vilone},\
  and\ \citenamefont {S\'anchez}}]{Vilone}%
  \BibitemOpen
  \bibfield  {author} {\bibinfo {author} {\bibfnamefont {A.}~\bibnamefont
  {Cardillo}}, \bibinfo {author} {\bibfnamefont {J.}~\bibnamefont
  {G\'omez-Garde{\~n}es}}, \bibinfo {author} {\bibfnamefont {D.}~\bibnamefont
  {Vilone}}, \ and\ \bibinfo {author} {\bibfnamefont {A.}~\bibnamefont
  {S\'anchez}},\ }\href {\doibase https://doi.org/10.1007/s11071-020-06021-7}
  {\bibfield  {journal} {\bibinfo  {journal} {New J. Phys.}\ }\textbf {\bibinfo
  {volume} {12}},\ \bibinfo {pages} {103034} (\bibinfo {year}
  {2010})}\BibitemShut {NoStop}%
\bibitem [{\citenamefont {Chung}\ \emph {et~al.}(2003)\citenamefont {Chung},
  \citenamefont {Lu},\ and\ \citenamefont {Vu}}]{SPECTRA}%
  \BibitemOpen
  \bibfield  {author} {\bibinfo {author} {\bibfnamefont {F.}~\bibnamefont
  {Chung}}, \bibinfo {author} {\bibfnamefont {L.}~\bibnamefont {Lu}}, \ and\
  \bibinfo {author} {\bibfnamefont {V.}~\bibnamefont {Vu}},\ }\href {\doibase
  10.1073/pnas.0937490100} {\bibfield  {journal} {\bibinfo  {journal} {Proc.
  Nat. Acad. Sci. USA}\ }\textbf {\bibinfo {volume} {100}},\ \bibinfo {pages}
  {6313} (\bibinfo {year} {2003})}\BibitemShut {NoStop}%
\bibitem [{\citenamefont {Restrepo}\ \emph {et~al.}(2007)\citenamefont
  {Restrepo}, \citenamefont {Ott},\ and\ \citenamefont {Hunt}}]{Restrepo}%
  \BibitemOpen
  \bibfield  {author} {\bibinfo {author} {\bibfnamefont {J.~G.}\ \bibnamefont
  {Restrepo}}, \bibinfo {author} {\bibfnamefont {E.}~\bibnamefont {Ott}}, \
  and\ \bibinfo {author} {\bibfnamefont {B.~R.}\ \bibnamefont {Hunt}},\ }\href
  {\doibase 10.1103/PhysRevE.76.056119} {\bibfield  {journal} {\bibinfo
  {journal} {Phys. Rev. E}\ }\textbf {\bibinfo {volume} {76}},\ \bibinfo
  {pages} {056119} (\bibinfo {year} {2007})}\BibitemShut {NoStop}%
\bibitem [{\citenamefont {Steinegger}\ \emph {et~al.}(2021)\citenamefont
  {Steinegger}, \citenamefont {Arola-Fern\'andez}, \citenamefont {Granell},
  \citenamefont {G\'omez-Garde{\~{n}}es},\ and\ \citenamefont
  {Arenas}}]{Benja2}%
  \BibitemOpen
  \bibfield  {author} {\bibinfo {author} {\bibfnamefont {B.}~\bibnamefont
  {Steinegger}}, \bibinfo {author} {\bibfnamefont {L.}~\bibnamefont
  {Arola-Fern\'andez}}, \bibinfo {author} {\bibfnamefont {C.}~\bibnamefont
  {Granell}}, \bibinfo {author} {\bibfnamefont {J.}~\bibnamefont
  {G\'omez-Garde{\~{n}}es}}, \ and\ \bibinfo {author} {\bibfnamefont
  {A.}~\bibnamefont {Arenas}},\ }\href@noop {} {\bibfield  {journal} {\bibinfo
  {journal} {In preparation}\ } (\bibinfo {year} {2021})}\BibitemShut {NoStop}%
\end{thebibliography}%

\end{document}